\documentclass{IEEEoj}
\usepackage{cite}
\usepackage{amsmath,amssymb,amsfonts}
\usepackage{algorithmic}
\usepackage{graphicx,color}
\usepackage{textcomp}
\def\BibTeX{{\rm B\kern-.05em{\sc i\kern-.025em b}\kern-.08em

    T\kern-.1667em\lower.7ex\hbox{E}\kern-.125emX}}
\AtBeginDocument{\definecolor{ojcolor}{cmyk}{0.93,0.59,0.15,0.02}}
\def\OJlogo{\vspace{-4pt}\hskip-4pt\includegraphics[height=18pt]{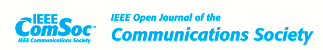}}

\let\labelindent\relax
\usepackage{enumitem}
\usepackage{cite}
\usepackage{amsmath,amssymb,amsfonts}
\usepackage{algorithmic}
\usepackage{enumitem}
\usepackage{graphicx}
\usepackage{textcomp}
\usepackage{xcolor}
\usepackage{csquotes}
\usepackage{balance}
\usepackage{url}
\usepackage{balance}
\usepackage{rotating}
\usepackage{tikz}

\def\BibTeX{{\rm B\kern-.05em{\sc i\kern-.025em b}\kern-.08em
    T\kern-.1667em\lower.7ex\hbox{E}\kern-.125emX}}

\usepackage{soul, color}
\usepackage[acronym,toc,shortcuts]{glossaries}
\usepackage{multirow}
\usepackage{comment}
\usepackage{soul}

\graphicspath{{figures/}}

\makeglossaries
\newacronym{3GPP}{3GPP}{The 3rd Generation Partnership Project }
\newacronym{5G}{5G}{Fifth Generation}
\newacronym{AAA}{AAA}{Authentication, Authorization and Accounting}
\newacronym{AI}{AI}{Artificial Intelligence}
\newacronym{ARIMA}{ARIMA}{AutoRegressive Integrated Moving Average}
\newacronym{BS}{BS}{Base Station}
\newacronym{CDN}{CDN}{Content Delivery Network}
\newacronym{COTS}{COTS}{Commercial Off-The-Shelf}
\newacronym{CPU}{CPU}{Central Processing Unit}
\newacronym{CU}{CU}{Centralized Unit}
\newacronym{DU}{DU}{Distributed Unit}
\newacronym{GEO}{GEO}{Geostationary Orbit}
\newacronym{gNB}{gNB}{Next Generation Node B}
\newacronym{GRU}{GRU}{Gated Recurrent Unit}
\newacronym{ISL}{ISL}{Inter-Satellite Link}
\newacronym{LEO}{LEO}{Low Earth Orbit}
\newacronym{MEO}{MEO}{Medium Earth Orbit}
\newacronym{ML}{ML}{Machine Learning}
\newacronym{NN}{NN}{Neural Network}
\newacronym{O-RAN}{O-RAN}{Open Radio Access Network}
\newacronym{QoS}{QoS}{Quality of Service}
\newacronym{PRB}{PRB}{Physical Resource Block}
\newacronym{xApp}{xApp}{eXtended application}
\newacronym{SFF}{SFF}{Simple-Feed-Forward}
\newacronym{LSTM}{LSTM}{Long-Short Term Memory}
\newacronym{OISL}{OISL}{Optical Inter-Satellite Link}
\newacronym{SBA}{SBA}{Service Based Architecture }
\newacronym{SN}{SN}{Seasonal-Naive}
\newacronym{MLP}{MLP}{ Multi-Layer Perceptron}
\newacronym{RNN}{RNN}{Recurrent Neural Network}
\newacronym{MSE}{MSE}{Mean Square Error}
\newacronym{MAE}{MAE}{Mean Absolute Scaled Error}
\newacronym{MAPE}{MAPE}{Mean Absolute Percentage Error}
\newacronym{ND}{ND}{Normalized Deviation}
\newacronym{NTN}{NTN}{Non-Terrestrial Network}
\newacronym{QL}{QL}{Quantile Loss}
\newacronym{RAN}{RAN}{Radio Access Network}
\newacronym{rApp}{rApp}{radio App}
\newacronym{RIC}{RIC}{RAN Intelligent Controller}
\newacronym{RRC}{RRC}{Radio Resource Control}
\newacronym{RTT}{RTT}{Round Trip Time}
\newacronym{RU}{RU}{Radio Unit}
\newacronym{SEC}{SEC}{Satellite Edge Computing}
\newacronym{SLAs}{SLAs}{Service Level Agreements}
\newacronym{SMO}{SMO}{Service Management and Orchestration}
\newacronym{UE}{UE}{User Equipment}
\newacronym{UPF}{UPF}{User Plane Function}

\begin{document}
\receiveddate{XX Month, XXXX}
\reviseddate{XX Month, XXXX}
\accepteddate{XX Month, XXXX}
\publisheddate{XX Month, XXXX}
\currentdate{11 January, 2024}
\doiinfo{OJCOMS.2024.011100}

\title{On the Architectural Split and Radio Intelligence Controller Placement in Integrated O-RAN-enabled Non-Terrestrial Networks}

\author{Jorge Baranda\IEEEauthorrefmark{1}\IEEEmembership{(Senior Member, IEEE)}, Marius Caus\IEEEauthorrefmark{1} \IEEEmembership{(Senior Member, IEEE)}, Luis Blanco\IEEEauthorrefmark{1}, Cristian J. Vaca-Rubio\IEEEauthorrefmark{1}, Engin Zeydan\IEEEauthorrefmark{1} \IEEEmembership{(Senior Member, IEEE)}, Kapal Dev\IEEEauthorrefmark{2}\IEEEmembership{(Senior Member, IEEE)}, Zheng Li\IEEEauthorrefmark{3}, Tomaso DeCola\IEEEauthorrefmark{4}}
\affil{Centre Tecnològic de Telecomunicacions de Catalunya (CTTC),  Spain}
\affil{Munster Technological University,Ireland}
\affil{Orange Innovation, France}
\affil{German Aerospace Center (DLR), Institute of Communications and Navigation, Germany}
\corresp{CORRESPONDING AUTHOR: Marius Caus (e-mail: marius.caus@ cttc.es).}
\authornote{This work was supported by the European Commission under the “5G-STARDUST” Project, which has received funding from the Smart Networks and Services Joint Undertaking (SNS JU) under the European Union's Horizon Europe research and innovation program under Grant Agreement No 101096573 and by UNITY-6G project, funded from European Union’s Horizon Europe Smart Networks and Services Joint Undertaking (SNS JU) research and innovation programme under the Grant Agreement No 101192650. 
Views and opinions expressed are however those of the authors only and do not necessarily reflect those of the European Union. Neither the European Union nor the granting authority can be held responsible for them. 
This work has received funding from the Swiss State Secretariat for Education, Research, and Innovation (SERI).}
\markboth{Preparation of Papers for IEEE OPEN JOURNALS}{Caus \textit{et al.}}

\begin{abstract}

The integration of Terrestrial Networks (TNs) with Non-Terrestrial Networks (NTNs) poses unique architectural and functional challenges due to heterogeneous propagation conditions, dynamic topologies and limited on-board processing capabilities. This paper examines architectural and functional split strategies that are consistent with O-RAN principles for future integrated TN-NTN systems. A taxonomy of split options is proposed that distributes RAN and core functions between satellites and ground nodes, and trade-offs in terms of performance, latency, autonomy and deployment are analysed. In particular, we evaluate configurations ranging from pure on-board DU deployments to full gNB and UPF integration into satellites, including variations based on intra- and inter-satellite processing. In addition, the placement of Near-RT and Non-RT RAN Intelligent Controllers (RICs) is discussed, proposing flexible split strategies between space and ground to optimise the performance and scalability of the control loop. A comprehensive mapping between architectural splits and RIC placement options is provided, emphasising implementation constraints and interoperability considerations. The paper concludes by identifying key challenges and outlining future directions to enable standardised, modular and efficient TN-NTN convergence in the context of the O-RAN.

\end{abstract}


\maketitle


\section{Introduction}
\label{intro}




\glspl{NTN} are emerging as critical infrastructure in the evolution of global wireless communication \cite{ntn_cioni, Corici_25}. Unlike traditional terrestrial networks, \glspl{NTN} can deliver connectivity in remote, rural, maritime, or disaster-stricken regions where ground infrastructure is impractical or economically unfeasible. Standardized by 3GPP in Release 17 and enhanced in Release 18, \ac{NTN} technologies now support 5G New Radio (NR) protocols with modifications for high latency, Doppler effects, and mobility patterns unique to non-terrestrial links. Release 19 is currently considering regenerative payloads. However, integrating this new paradigm into reliable, scalable communication systems introduces significant technical challenges. These include propagation delay, particularly in use-cases based on \ac{GEO} satellites, rapid handovers and beam tracking in \ac{LEO}  constellations, limited onboard processing capabilities, and the need for efficient spectrum utilization across dynamically changing topologies.

Despite rapid advancements in satellite networks, most NTNs remain largely decoupled from terrestrial orchestration frameworks. Their control-plane architectures are typically monolithic and statically configured, which limits their ability to respond to dynamic traffic demands, shifting topologies, and evolving mission objectives. As a result, these systems often suffer from inefficient spectrum allocation and chronically underutilized network capacity. While \ac{AI} offers great potential for autonomous decision-making and adaptive resource management, its practical deployment onboard spaceborne platforms is constrained by limitations in the computational power, energy availability, and intermittent connectivity.

Recent advances in NTN payload capabilities, such as regenerative satellites and digitally steerable beamforming arrays, have increased the feasibility of deploying intelligent radio access functions directly in spaceborne platforms. In this context, the \ac{O-RAN} architecture \cite{Polese23} —originally developed to disaggregate and virtualize terrestrial RAN functions— has emerged recently as a key enabler for managing Non-Terrestrial Networks in 6G  \cite{Campana23, Mahboob25}. O-RAN's open interfaces, software-defined components, and AI-driven control via RAN Intelligent Controllers (RICs) make it highly adaptable to the constraints and requirements of NTN deployments \cite{oran2025ntn}. The disaggregated architecture allows flexible placement of \glspl{CU}, \glspl{DU}, and \glspl{RU} across satellite and ground segments, while AI-enabled orchestration supports dynamic resource allocation, beam management, and energy-efficient scheduling under latency and power constraints.

This paper presents an extensive analysis of various architectural splits to enable the adoption of O-RAN-based architecture in non-terrestrial networks, focusing on the specific challenges and constraints inherent to satellite and aerial platforms. It systematically evaluates the advantages and disadvantages of each split option in the context of NTNs, providing insights into their feasibility, performance, and implementation complexity and open challenges. Furthermore, the paper conducts a thorough investigation into the placement of the \ac{RIC} components within O-RAN-based NTNs, analyzing the trade-offs, challenges and network scalability. Through this analysis, the papers offers a comprehensive understanding of the key design considerations for realizing efficient and robust O-RAN integration in NTNs.

\section{Related Work and Main Contributions}




The paper \cite{Campana23} did a preliminary exploration of integrating \ac{O-RAN} with \glspl{NTN}, especially satellite systems, to support future 5G/6G connectivity. It highlights how O-RAN's modular, intelligent architecture can optimize NTN performance through AI-driven RICs.  F. Veisi et al. discussed in \cite{Veisi25} on the integration of disaggregated RAN (dRAN) with \glspl{NTN} like satellites to enhance 5G/6G scalability and connectivity. It highlights challenges such as link reliability and satellite limitations. The paper explores deployment architectures and proposes a hybrid Free-Space Optical/Radio Frequency (FSO/RF) communication approach to improve reliability and service consistency in NTN-dRAN systems.

Reference \cite{Mahboob25} explores how integrating Open Radio Access Network (O-RAN) architecture with NTN can transform future 6G mobile networks. It highlights O-RAN’s modular, AI-driven, and open interfaces as ideal for tackling NTN challenges like high latency, mobility, and limited onboard resources. The authors propose and analyze several integrated architectures—including transparent, regenerative, and regenerative-DU payloads. A case study demonstrates that deploying the near-real-time RIC on the ground, combined with deep learning-based prediction of delayed satellite feedback, can closely match the performance of satellite-based RICs. The paper concludes with future research directions in AI-driven optimization, energy efficiency, and advanced computing to support O-RAN-NTN integration in 6G.

The work in \cite{Daurembekova24} investigates the dynamic placement of \ac{RAN} functions in such unified 3D architectures. Although it is not strictly relying on O-RAN-based architectures, it puts emphasis on the complexities of \ac{DU} interactions with \glspl{RU} and \glspl{CU}. The analysis offers strategic insights into optimizing RAN deployment to ensure seamless, high-performance operation in next-generation networks. The authors of \cite{baena2025space} present Space-O-RAN, a novel distributed control architecture that extends Open RAN into satellite constellations through hierarchical, closed-loop control. Lightweight dApps onboard satellites enable real-time tasks like scheduling and beam steering without continuous ground contact. Simulations on a Starlink-like \ac{LEO} network confirm feasible latency for both real-time and strategic operations. The architecture enables scalable, intelligent, and interoperable non-terrestrial networks for 6G systems. Recently, O-RAN alliance published a white paper \cite{oran2025ntn} that explores how O-RAN's open and intelligent RAN architecture can be integrated with 3GPP-based NTNs, supporting both transparent and regenerative satellite payloads across various orbits. It highlights use-cases such as beam management and RIC-enhanced analytics. The paper reviews relevant O-RAN and 3GPP standardization efforts and ongoing work. Security aspects like feeder link protection and function split selection are addressed. Overall, it provides a high-level overview of deploying O-RAN in satellite-based networks to enable global connectivity.

Reference \cite{Lee_25} proposes an open NTN architecture based on the O-RAN model, enabling multiple service operators to share LEO satellites cost-effectively. The architecture places \glspl{CU} on satellite ground stations, while \glspl{DU} and \glspl{RU} are onboard the satellites. The key challenge addressed is the limited bandwidth of \ac{ISL} fronthaul, unlike the wired fronthaul in terrestrial networks. The authors formulate a joint optimization problem involving compression rates and power allocation for data and pilot signals, aiming to maximize uplink sum rates under ISL constraints. They solve it using an alternating optimization approach. The work in \cite{Probabilistic_25} deals AI-enhanced resource management in O-RAN based NTN. It explores the limitations of traditional single-point prediction models, which often fall short due to the complexity of satellite dynamics and variable signal conditions. Instead, it evaluates probabilistic forecasting methods \cite{kasuluru2025enhancing} for predicting bandwidth and capacity needs. Results highlight the superior performance and uncertainty quantification capabilities of these models, making them well-suited for optimizing resource allocation in O-RAN-based NTN systems.


Current specifications for LEO satellite networks exhibit significant technical constraints that make difficult scalable, autonomous operation. Constellation-wide management remains limited, with only basic adaptations of Self-Organizing Networks (SON) and lacking support for distributed scheduling, in-orbit resource orchestration, or AI-driven control under dynamic network conditions. LEO satellites require efficient resource usage. Resource demand varies by time and location, influenced by population density and economic activity of the illuminated area, requiring adaptive resource allocation. 
Current autonomous RAN capabilities are restricted to narrow use cases, such as store-and-forward for IoT, and fail to ensure continuity without persistent ground access or in the presence of multi-hop inter-satellite links. Centralized frameworks are poorly suited to LEO environments due to assumptions of stable control-plane connectivity or high latency tolerance, and limited support for compute-aware resource allocation. Furthermore, AI integration in NTN faces challenges from latency variability, distributed and sparse datasets, and limited onboard computing resources, undermining fine-grained inference tasks. Similarly, emerging RAN architectures using RAN Intelligent Controllers assume presume static backhaul, which is misaligned with the volatile nature of next generation NTNs. In summary, while standardization efforts are progressing, current frameworks fall short of enabling the decentralized, adaptive control needed for dynamic and resilient satellite constellations.

This paper investigates how O-RAN architecture can be extended and tailored specifically for NTN scenarios. By focusing on satellite-native implementations of O-RAN, we explore design strategies for architectural splits, latency-tolerant interfaces, and distributed intelligence across heterogeneous altitudes. We present reference architecture for O-RAN-enabled NTNs, propose enhancements to RIC-driven control loops suited for space-based operation, and analyze key protocol adaptations needed to support mobility, link intermittency, and onboard processing limits. Our aim is to establish a foundational approach for future NTN systems that are open, modular, and intelligent by design, capable of meeting the coverage, resilience, and performance requirements of 6G and beyond. In particular the key contribution of this paper are the following: 

\begin{itemize}
    \item \textit{Taxonomy of architectural O-RAN splits for NTN:} Proposes a detailed taxonomy of architectural and functional split options in line with O-RAN principles based on the available capabilities in the spaceborne segment.
    
    \item \textit{Systematic evaluation of split configurations:} Analyzes a wide range of architectural split scenarios for O-RAN NTN, from ground-based processing to fully on-board deployments (e.g., gNB and UPF on satellites), including intra- and inter-satellite processing strategies.
    
    \item \textit{Trade-off analysis:} Provides a comprehensive analysis of the trade-offs associated with different architectural splits in terms of performance, latency, autonomy, scalability, and deployment complexity.
    
    \item \textit{RIC placement strategies:} Investigates optimal placement strategies for Near-RT and Non-RT RAN Intelligent Controllers in space-ground split environments, addressing control loop efficiency, computational requirements, robustness to intermittent feeder link conditions, persistent ground access, scalability and open challenges (e.g., the need for new interfaces, limitations, requirements, etc).
    
    \item \textit{Mapping of architectural splits and RIC placement:} Delivers a novel mapping between architectural split options and RIC deployment alternatives, emphasizing implementation feasibility and interoperability within O-RAN-based NTNs.
    
    \item \textit{Design guidelines for O-RAN in NTNs:} Offers practical insights and design considerations for implementing modular, efficient, and standardized O-RAN architectures in satellite and aerial networks.
    
    \item \textit{Identification of NTN-specific challenges and future research directions:} Highlights the unique constraints of NTNs (e.g., dynamic topologies, limited onboard processing, heterogeneous links) that impact O-RAN integration and architectural design. Furthermore, the paper outlines open challenges and future research areas necessary to advance TN-NTN convergence and ensure effective O-RAN adoption in non-terrestrial contexts.
\end{itemize}

The rest of the paper is organized as follows. Section \ref{sec:background} provides technical context on O-RAN architecture and the challenges of applying it to NTNs.  Section \ref{sec:splits}  introduces different architectural options for distributing gNB functions in space-based networks. Section \ref{sec:rics} examines various strategies for distributing RIC functions across terrestrial and satellite nodes. Section \ref{sec:discussion} highlights challenges and proposes future research directions. Finally, Section \ref{sec:conclusions} concludes the paper by summarizing the proposed architectural options and highlighting their potential impact on NTN systems.

\section{Background}
\label{sec:background}



\subsection{O-RAN Architecture Overview} 
\label{sec:oran_overview}

5G has opened the door for a revolution in the architecture and management of mobile networks. With the introduction of the \ac{SBA} concept at the mobile core, mobile networks are leaving behind closed architectures. Mobile network components are no longer monolithic blocks employing dedicated protocols for direct communications among entities. This trend is also followed in the \ac{RAN} segment through the
propositions of the Open RAN (O-RAN) alliance \cite{o-ran-alliance}. It promotes openness, intelligence, and interoperability by the disaggregation of traditional monolithic base station architecture into flexible, virtualized components connected by standardized interfaces and naturally including close-loop operations enhancing network performance. In this way, the proposed model fosters vendor-neutral deployments, avoiding lock-in situations to operators, thus improving cost-efficiency and flexibility to adapt network deployments to meet the requirements of challenging use cases proposed by vertical industries.  More specifically, the foundations of O-RAN are based on four pillars:

\begin{itemize}[leftmargin=*]
\item{\textit{Dissagregation}}: the \ac{gNB} is decoupled into three logical units, namely \ac{RU}, \ac{DU}, and \ac{CU}. This allows operators to upgrade or replace hardware and software independently. 
\item{\textit{Open Interfaces}}: all functional splits between entities (e.g., between the \ac{RU}, \ac{DU}, and \ac{CU}) are standardized, enabling multi-vendor deployments.
\item{\textit{Cloud-Native Virtualization}}: O-RAN enables RAN functions to run on \ac{COTS} hardware, leveraging network function virtualisation (NFV) and containers.
\item{\textit{Intelligence and Automation}}: O-RAN proposes the use of \glspl{RIC} to support AI/ML-driven optimization and dynamic control at different timescales. 
\end{itemize}

Based on these benefits, it is reasonable that in future 6G networks, chasing interoperability between NTN and TN segments, the NTN components shift towards such O-RAN paradigm to provide the required flexibility in terms of network management and optimization. Next subsections, will present the main entities and interfaces proposed by current O-RAN specification to later sketch the potential challenges which motivates the architectural discussion in next sections.

\begin{figure*}[htbp!]
	\centering
    \colorbox{white}{\includegraphics[width=0.8\textwidth] {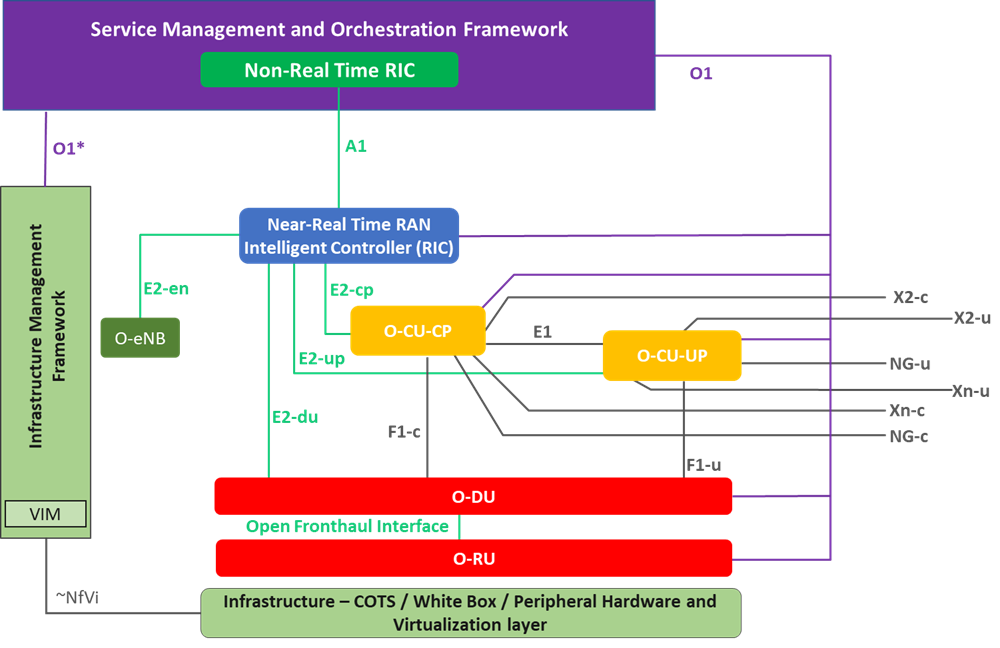}}
	\caption{Generic O-RAN architecture with interfaces between functional and intelligence components~\cite{o-ran-alliance}}\label{fig:oran_arch}
\end{figure*}

\subsubsection{O-RAN Functional Entities and Interfaces}
Figure \ref{fig:oran_arch} presents the RAN architecture proposed by O-RAN Alliance specifications~\cite{o-ran-alliance}. The first thing to notice is the disaggregation of the gNB into three different network functions, as previously mentioned, thus allowing network management multiplexing gains due to distributed deployments.  These functions are: 
\begin{itemize}[leftmargin=*]
    \item \textit{O-RU:} manages the radio front-end and lower PHY layer operations (e.g., FFT/IFFT, beamforming) and transmits signals to the \glspl{UE}.
    \item \textit{O-DU}: handles real-time lower-layer protocol processing functions (e.g., Layer 1/2 MAC, RLC), such scrambling, modulation, and resource element mapping, close to the O-RU.
    \item \textit{O-CU:} Executes non-real-time higher-layer protocol processing function (e.g., PDCP, SDAP) which are further separated into control and user plane functions, often deployed in centralized data centres.
\end{itemize}
At the Intelligence layer, the O-RAN architecture proposes the use of RICs, namely:
\begin{itemize}[leftmargin=*]
    \item \textit{Near-Real Time (RT) RIC:} Offers control functions such as load balancing, mobility management, and interference mitigation in near-real-time (10 ms–1s). This is done through the use of xApps, which are custom logic applications responsible for collecting information, computing it, and triggering control actions to the O-RAN network functions. 
    \item \textit{Non-RT RIC:} policy-based guidance, analytics, ML model management collaborating with the the Near-RT RIC in control-loops bigger than 1s. It hosts rApps application supporting RAN optimization and management tasks. It is placed within the \ac{SMO} Framework 
\end{itemize}
Finally, we have the \ac{SMO} Framework, provides the end-to-end lifecycle management (i.e., orchestration and FCAPS (Fault, Configuration, Accounting, Performance and Security) operations) of the O-RAN network functions and RICs. Regarding the interfaces, O-RAN specifications include the following ones: \begin{itemize}[leftmargin=*]
    \item \textit{Open Fronthaul Interface (OFH):} based on the functional split 7.2x proposed by 3GPP, designed to  strike a balance between the simplicity of the O-RU and the data rates and latency requirements on the interface between the O-DU and O-RU.
    \item \textit{F1:} defined by 3GPP, connects the O-CU and O-DU to exchange control and user plane traffic.
    \item \textit{E2 Interface:} Enables the Near-RT RIC to collect data from the O-RAN network functions (i.e., O-CU and O-DU) and provide actions.
    \item \textit{E1 Interface:} defined by 3GPP, E1 
    enables independent scaling and management of the control and user plane functions within the O-CU.
    \item \textit{O1 interface:} Enables the \ac{SMO} to perform FCAPS of O-RAN network functions (excluding O-RU) and the Near-RT-RIC.
    \item \textit{A1 interface:} employed by the Non-RT RIC to provide policies, enrichment information and ML models to the Near-RT RIC and collect feedback.
\end{itemize} 

It is important to remark that the RAN disaggregation concept cannot be straightforwardly applied to NTN. The limitation essentially comes from the bit rate and the latency requirements of fronthaul and midhaul networks, which respectively connect RU-DU and DU-CU. Some details a provided hereinafter following the methodology proposed in \cite{Lar19}. The downlink and uplink fronthaul bit rate is defined by
\begin{equation}
BR=M\times N \times L \times \text{BTW} \times 2/T_s +\text{CR}.
\end{equation}
The bit rate is based on  the bit-width (BTW) which is the number of I and Q bits, the number of subcarriers $M$, the number of symbols $N$, the number of layers $L$, the slot duration $T_s$ and the  control signaling rate (CR). The CR is determined by the overhead regarding beamforming, resource element mapping and scheduling. For 20 MHz bandwidth, 64-QAM and 2 layers, the resulting rate is CR=1.856 Mbps. The rate linearly scales with the bandwidth, the modulation order and the number of layers. Accordingly, the rate for different system parameters can be easily obtained from the reference value. The maximum latency for a fronthaul network is limited to 500 $\mu s$.   

The downlink and uplink midhaul bit rate is given by
\begin{equation}
\text{BR}=\text{PR}+\text{CR},    
\end{equation}
where PR denotes de peak rate. The control and signaling rate CR is calculated as 24 Mbps and 16 Mbps in downlink and uplink, respectively. For 20 MHz bandwidth, 16-QAM and 1 layer, the peak rate in uplink is PR=50 Mbps. In the downlink, for 20 MHz bandwidth, 64-QAM and 2 layers, the peak rate is PR=150 Mbps. The parameter CR is assumed constant, while the peak rate PR linearly scales with the bandwidth, the modulation order and the number of layers. Accordingly, the rate for different system parameters can be easily obtained from the reference value. The maximum latency for a midhaul network is between 1.5 and 10 ms. Since the latency is not limited by the HARQ procedure, 
thus the a midhaul segment tolerates high latency. The bit rate requirements are computed in Table \ref{table_req}. As for the air interface, the following assumptions have been made: i) the cell is fully loaded, ii) the bit rate is computed in the downlink, iii) there are 14 OFDM symbols per slot, iv) IQ samples are represented with 14 bits, v) there is one layer and one antenna port and vi) modulation order up to 64-QAM.

\begin{table}
\centering
\caption{Bit rate and latency requirements for different transmission bandwidth}
	\begin{tabular}{|l|c|c|c|}
	\hline
	        Bandwidth  (MHz)                   & 100 & 200 & 400\\ \hline
	  Subcarrier spacing (KHz) & 60 &  60 & 120 \\ \hline
	   Number of PRBs & 132	 & 264	 & 264	\\ \hline
	   Fronthaul bit rate (Gbps)  & 2.48 & 4.97 & 9.96\\ \hline
	   Midhaul bit rate (Gbps)	 & 399 & 774 & 1.524 \\ \hline
	\end{tabular}
\label{table_req}
\vspace{-.5cm}
\end{table}

As expected, the fronthaul network demands more resources than the midhaul network. The reason lies in the fact that the interface shall transport samples. Conversely, the midhaul network calls for more relax bit rate requirements. 

\subsection{Considerations and Challenges for extending O-RAN to NTNs}
NTN, focusing on LEO/MEO/GEO satellites,  which is the object of this paper, are integral in 6G networks to provide global and ubiquitous coverage, as previously mentioned. The different characteristics on the different orbits, requires different consideration to adopt and propose adaptations to the canonical O-RAN architecture (e.g., inclusion of dApps or spaceApps (\cite{baena2025space}) and the split of the different O-RAN network functions between the satellite and ground station locations. Such splits are discussed in the following sections taking into account the following challenges:
\begin{itemize}[leftmargin=*]
    \item \textit{Latency and Timing Constraints:} NTN links introduce propagation delays (e.g., 5–20 ms for LEOs, larger than 250 ms for GEOs), which have impact on control-plane signaling and real-time interfaces like E2 and F1. Buffering, synchronization, and jitter handling mechanisms must be adapted.
    \item \textit{Doppler and Link Variability:} Satellite movement leads to Doppler shifts and dynamic link changes (e.g., unstable performance of feeder link), which are not considered in current O-RAN PHY/MAC-level standards. This demands for enhanced PHY designs and scheduler algorithms. 
    \item \textit{Interface and Protocol adaption:} O-RAN interfaces assume consistent, low-latency terrestrial links. In NTN, protocols must be enhanced with mechanisms for delay tolerance, retransmission, and session persistence. In addition to this, for Non-Geostationary Orbit (NGSO), visibility with the gateway (GW) on ground can change, thus introducing additional routing considerations to exchange data between satellites running O-DUs  and O-CUs and near-RT RICs. Indeed, for NGSO orbits, the continuous movement of the \textit{gNB} introduces an additional dimension to the handover problem and its connections towards entities running on ground. It also advocates for distributed and synchronised architectures of near-RT RICs and non-RT RICs. 
    \item \textit{Interoperability with TNs}: TNs and NTNs must share spectrum, mobility anchors, and user session management. This requires tight integration between 3GPP Core networks, \ac{SMO}, and both RICs to support dynamic routing and resource orchestration.
    \item \textit{Security and Reliability Risks}: the openness of O-RAN interfaces, software-based deployments and distribution of the network functions in regenerative satellite payload expands the attack surface raising concerns around encryption, jamming, and system integrity. This can be extremely critical for defense-communications.
    \item \textit{Hardware, computational constraints and CAPEX}: Although advancements in the expenses associated with the deployment of satellite constellations, such deployments are still expensive and require an accurate preparation. Moreover, satellite operates under strict power, processing capacity and thermal management conditions, thus introducing fluctuations in the operations they can handle, including possible reconfiguration of network functions, and algorithms they can host and run. 
\end{itemize}


\section{Proposed Architectural Split Options in Integrated NTN Systems}
\label{sec:splits}

When integrating \glspl{NTN} into 5G architectures, the deployment of gNB components in the space and terrestrial segments is crucial. In the following, we discuss three key options for functional split, with increasing levels of satellite autonomy and complexity.

\subsection{Option 1: DU and RU On Space (Split 2)}
\label{subsec:du_space}


In this approach, the \ac{DU} of the gNB is deployed onboard the satellite, while the \ac{CU} remains on the ground. This corresponds to a Split 2 architecture in the 5G RAN, in which the satellite handles physical layer functions such as modulation, channel coding and real-time scheduling, as well as procedures such as the random access, the hybrid automatic repeat request (HARQ) retransmission scheme and the system information block 19 (SIB19) generation. The main advantage of this setup is the reduction of latency for time-sensitive PHY/MAC functions. However, there are also limitations with this architecture:
\begin{itemize}[leftmargin=*]
    \item The CU remains on Earth, limiting responsiveness and dependence on terrestrial links.
    \item Application-layer services and edge computing capabilities are not feasible onboard.
\end{itemize}

\begin{figure}[htp!]
	\centering
    \includegraphics[width=0.45\textwidth] {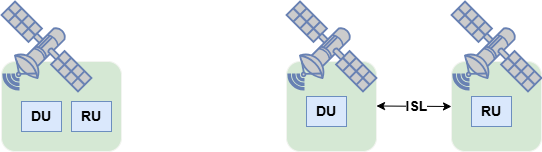}
	\caption{Deployment options for DU in space. Left: DU in single satellite. Right: DU and RU interfaced over ISL.}\label{fig:option1}
\end{figure}

Regarding inter-satellite mobility, when a UE moves from the source to the target satellite, two type of procedures can be differentiated. More precisely, mobility is handled as either intra- or inter-CU handover, depending on whether the source and target satellites are connected to the same CU or to different CUs, respectively. A specific characteristic of the intra-CU handover is that it does not require involvement of the core network, as shown in \cite{nr401}.  The schematic view of the architecture is depicted in Figure \ref{fig:option1}. Two topology options are presented, which are described hereinafter. 

\subsubsection{Option 1a: DU in single satellite}
In this configuration the satellites of the constellation are identical, each hosting the DU and the RU. Moreover, each satellite must establish feeder links to connect the on-board DU with the CU, which is co-located to the satellite gateway. Therefore, the F1 interface shall be implemented over the feeder link. It is important to remark that the F1 interface requires a persistent connection. For non-geostationary orbit (NGSO) constellations, this means  that satellites must be able to maintain two feeder links, while the connection is transferred from the source to the target gateway. The support of variable latency and seamless handovers becomes the primary challenges of the feeder link.  

From the signaling point of view, it is worth highlighting that the \ac{RRC} would reside in the CU. Hence, CU-DU interactions over F1 interface will include the procedures detailed in \cite{nr470}. Consequently, signaling requirements must be carefully considered to ensure proper dimensioning of the feeder link. During a feeder link switchover, the DU should handover all the UEs under the satellite coverage.  Remarkably, this procedure requires a very high signaling load.  



\subsubsection{Option 1b: DU and RU in different satellites}
The option 1b offers a more distributed architecture in which satellites are not identical. In this regard, there are two classes of satellites to enable RAN disaggregation. Class 1 satellites are equipped with a DU and include feeder links. The RUs are implemented in the class 2 satellites. The fronthaul interface that connects the DU and the RU is implemented over an inter-satellite link (ISL). In this deployment, class 1 satellites include four permanent ISLs, two connecting adjacent satellites in the same orbital plane and two connecting to adjacent orbital planes. Therefore, the DU is connected to 4 RUs, which are located in different satellites. 

The primary technical challenge of this deployment arises from the \ac{RTT} delay between adjacent satellites, which is significantly larger than typical RU-DU delays encountered in terrestrial networks. Remarkably, adjacent satellite could be separated hundreds of kilometers, or even up to a thousand kilometers. The dominant technology in terrestrial deployments to interface the DU and the RU is the common public radio interface (CPRI). However, the specification reported in \cite{cpri} highlights that this technology is not suitable to transport digitized signals over ISLs. The limitation is that the maximum one-way delay supported by eCPRI is 500$\mu s$. The RTT delay in most ISLs exceeds the maximum value specified by eCPRI. This observation highlights the necessity of using a technology specifically tailored to NTN. 

\subsection{Option 2: Full gNB In Space}
\label{subsec:full_gnb}

In this configuration, the complete gNB stack comprising the RU, DU, and CU is deployed onboard the satellite. This setup eliminates the need for fronthaul and midhaul interfaces between terrestrial entities and reduces the dependence on Earth-based control and scheduling. It corresponds to a fully autonomous gNB entity \cite{nr-ntn, masini20235g} according to 3GPP NR architecture, with N2/N3 interfaces directly connecting the satellite to the 5G Core (5GC) through feeder links\cite{lin20215g}. By colocating the entire RAN protocol stack onboard \cite{rossato2024simulation}, the system executes the complete sequence of 5G RAN functions:
\begin{itemize}[leftmargin=*]
    \item The physical layer performs FFT/iFFT operations, LDPC/BCH encoding, cyclic prefix addition and beamforming.
    \item The MAC and RLC layers execute time-critical functions such as resource scheduling, HARQ retransmissions and radio bearer reconfiguration.
    \item The RRC and SDAP layers handle control signaling, mobility procedures, bearer establishment, and QoS mapping.
\end{itemize}
The onboard integration removes the stringent timing requirements of midhaul and fronthaul interfaces, which typically cannot be met in NTN environments due to propagation delays. 

A core benefit of this option lies in the simplification of RAN–Core Network (CN) integration. As the full gNB stack is onboard, the satellite only requires the N2 and N3 interfaces to connect with the 5G Core via feeder links, avoiding complex control synchronization with Earth-based CUs. This reduces signaling overhead and allows for fast UE context management\cite{tsegaye2024towards, masini20235g}, especially relevant for intra-satellite mobility and session continuity. Moreover, by embedding all scheduling and RRC procedures locally, this approach enhances autonomy and responsiveness, particularly in cases where gateway visibility is intermittent or disrupted. Additionally, mobility anchoring and bearer management are performed on orbit, which allows for seamless UE handovers at the gNB level, reducing the signaling burden on the CN. These advantages make Option 2 especially attractive for NTN systems requiring low-latency access and mobility robustness.

However, the absence of an onboard \ac{UPF} introduces important limitations. As the N6 interface is not terminated in space, no localized breakout or PDU-level processing is feasible. All user-plane traffic must be routed to the ground-based UPF, increasing the load on feeder links and introducing additional latency for application-level services. Consequently, Option 2 does not support caching, CDN integration, or Satellite Edge Computing (SEC) capabilities. This lack of edge functionality renders the architecture suboptimal for data-intensive or real-time applications that would benefit from local service exposure. Additionally, the integration of the full gNB stack onboard significantly increases power consumption and processing requirements. Satellites must be equipped with high-performance, onboard processing units capable of handling real-time PHY/MAC tasks, ciphering and control signaling, which may be infeasible for small satellite platforms with limited power budgets.

\begin{figure}[htp!]
	\centering
    \includegraphics[width=0.45\textwidth] {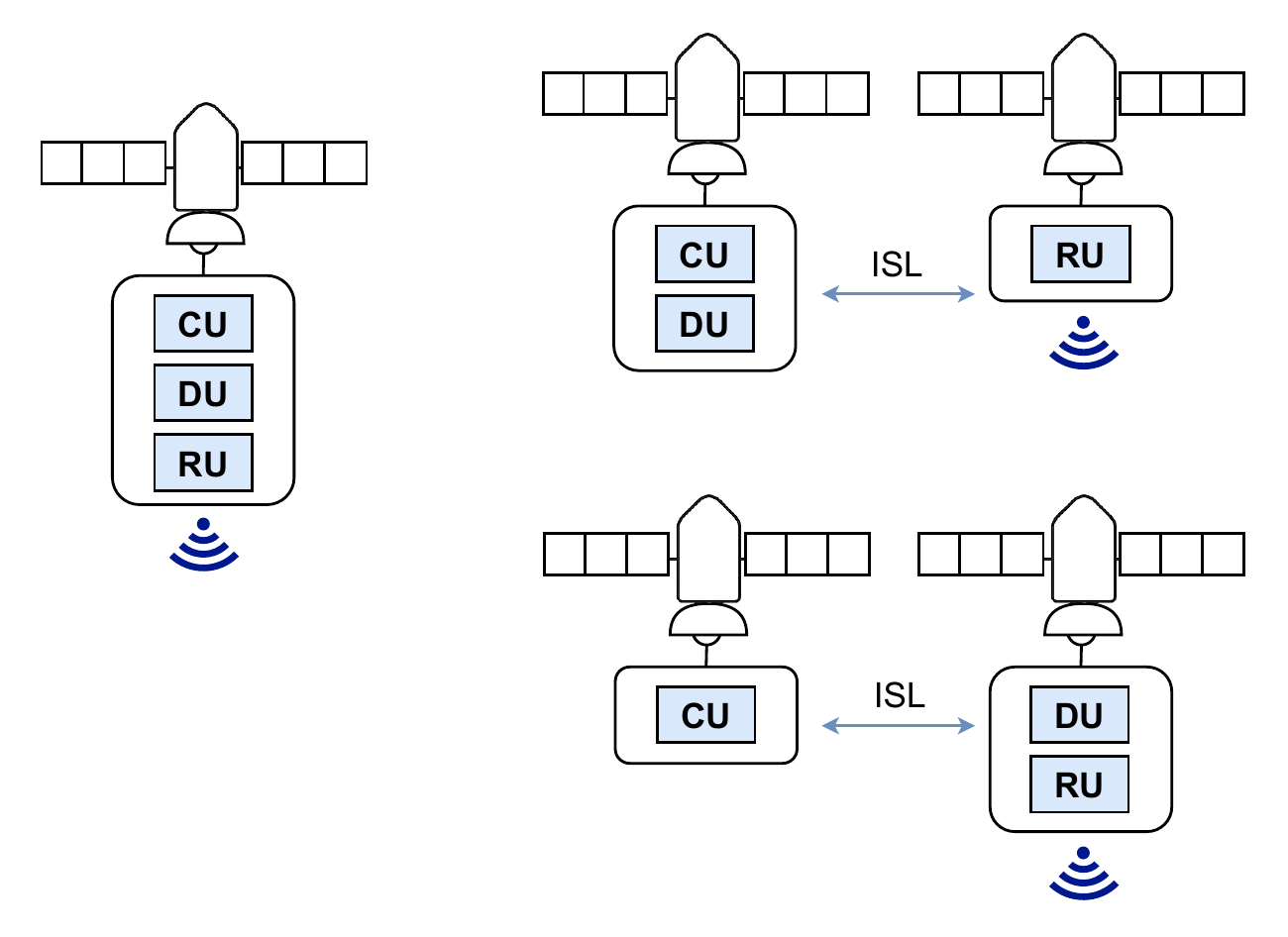}
	\caption{Deployment options for full gNB in space. Left: Option 2a (monolithic gNB). Right-top: Option 2b with RU separated over ISL. Right-bottom: Option 2b with DU and RU together and CU separated. ISLs support fronthaul or F1-C/U, depending on split.}\label{fig:option2}
\end{figure}

While this configuration reduces timing constraints on the feeder link, it still demands persistent, high-throughput N2/N3 connectivity to the core network. For NGSO constellations, the dynamic motion of satellites requires robust mechanisms for gateway reassignment and session re-anchoring to ensure continuity. A potential solution for managing these feeder link handovers is the NG-Flex configuration specified in 3GPP TS 38.410 \cite{3gpp38.410}. In this setup, each NG-RAN node can be connected to multiple AMF Sets within an AMF Region, as long as the corresponding slice is supported. This architecture, complemented by definitions in TS 23.501 \cite{3gpp23.501}, allows the onboard gNB to switch among ground-based AMFs without requiring session reestablishment, thus simplifying mobility anchoring across gateways. This makes NG-Flex particularly suitable for Option 2 deployments with full gNB in space, where feeder link transitions occur frequently due to satellite motion. Moreover, managing IP mobility, synchronizing bearer contexts and maintaining QoS guarantees under variable link conditions remain open research challenges. Furthermore, Option 2 provides an effective compromise between latency performance and architectural complexity, offering a high degree of autonomy and fast control-plane responsiveness while remaining compatible with current gNB/CN interface standards. Nevertheless, its lack of edge-service support and the elevated onboard processing requirements necessitate careful design trade-offs. It is best suited for medium-class LEO or MEO satellites capable of hosting the complete RAN stack but not yet supporting full edge computing. This configuration is particularly applicable to scenarios prioritizing mobile broadband connectivity, efficient mobility handling and simplified terrestrial integration without the additional complexity of local user-plane functions. To accommodate varying architectural constraints and mission objectives, two distinct deployment variants of the full on-board gNB can be considered: Option 2a implements a monolithic integration of the CU, DU, and RU within a single satellite platform, while Option 2b distributes these functional blocks across multiple satellites interconnected via ISL links, as illustrated in Fig. \ref{fig:option2}.

\subsubsection{Option 2a: Monolithic gNB}
In this variant, the entire gNB stack is fully embedded within a single satellite platform. This monolithic integration simplifies internal communication and removes the need for any inter-satellite synchronization or high-speed link coordination. All intra-gNB interfaces (e.g., fronthaul, midhaul) become local, effectively eliminating latency variability caused by long-distance physical separation between components. This is particularly beneficial for timing-sensitive processes like HARQ feedback, MAC scheduling, and Random Access Channel (RACH) response handling. Since all functions share the same physical platform, coordination is inherently tight, and software integration can leverage shared memory or fast bus-level interconnects, achieving minimal processing delay.

This approach is the most straightforward to implement from both a protocol and hardware standpoint. It reduces system complexity, limits the risk of synchronization errors, and facilitates thermal co-design of processing units. Furthermore, it simplifies RAN management and orchestration as the entire node can be managed as a single logical and physical entity. Deployment, provisioning, and fault recovery operations are also more contained and deterministic. Because intra-satellite delays are negligible (on the order of microseconds), real-time control loops such as CSI reporting, scheduling decisions and HARQ processes can function with latencies comparable to those in terrestrial networks. However, this configuration imposes a non-trivial burden on the satellite onboard processing unit (OBP), which must support the entire gNB protocol stack, including high-rate baseband signal processing and real-time scheduling decisions. Depending on the supported bandwidth, number of antenna ports and expected user load, the satellite must be provisioned with high-performance CPUs, possibly accompanied by FPGAs or GPUs for acceleration of PHY layer operations. The resulting power, mass, and thermal requirements may exceed the capabilities of small LEO platforms, suggesting this architecture is more suitable for medium or large LEO/MEO satellites with adequate payload capacity.

Another operational consideration is that the satellite must maintain continuous or quasi-continuous feeder link visibility with at least one gateway, as both the N2 and N3 interfaces require persistent connectivity to the 5G Core. Gateway handovers must be managed carefully to prevent session interruption, requiring dual gateway tracking or fast reattachment procedures. Nonetheless, the full co-location of the gNB stack allows these handovers to be more manageable compared to distributed setups, as there is no dependency on external RAN functions that could become unreachable during a feeder link handover. Consequently, Option 2a represents a baseline configuration for autonomous NTN RAN nodes and serves as a feasible intermediate step toward fully edge-enabled satellites in Option 3. Its simplicity and robustness make it an attractive choice for mission profiles where processing integration is preferred over architectural distribution and where multi-satellite coordination introduces unacceptable risk or complexity.
\subsubsection{Option 2b: Distributed gnB in space}
In contrast, Option 2b explores a distributed approach in which the components of the gNB stack, specifically the RU, DU, and CU, are spread across different satellites. For example, one satellite might host the CU and DU, while another, physically separated satellite hosts the RU. These components would then be connected via \glspl{ISL}, forming a logical gNB node across a spatially distributed platform. The primary motivation for this approach is architectural scalability: by decoupling the RU and DU/CU roles across satellites, one can imagine a constellation where certain satellites specialize in radio front-end operations (analog/RF + lower PHY), while others aggregate digital baseband processing and control functions.

This modular separation allows for potential reusability of satellite types and dynamic reassignment of RAN functions across the constellation. For example, a group of RU satellites could beamform toward different ground coverage zones while a more powerful DU/CU satellite manages their scheduling and mobility procedures in a centralized manner. This supports flexible coverage shaping and resource pooling, especially in large constellations.

However, this distribution brings significant challenges. The most critical is the added delay and synchronization overhead introduced by the inter-satellite links. Even with advanced ISLs operating at Gbps speeds and microsecond latencies under ideal conditions, the physical separation between satellites of typically hundreds of kilometers, results in non-negligible one-way delays. Thus, the standard fronthaul protocol stack is not directly applicable in this scenario, requiring custom transport protocols or enhanced synchronization techniques capable of compensating for larger propagation delays and jitter.

In addition, the physical mobility of satellites in low Earth orbit results in constantly changing inter-satellite topologies. Maintaining reliable ISL connectivity requires active routing and dynamic link-state awareness. When multiple RU satellites feed into a shared DU/CU satellite, load balancing, link failures, and variable traffic patterns must be handled in real time, necessitating advanced inter-satellite control-plane signaling and possibly distributed RIC (RAN Intelligent Controller) instances. The need for synchronization of HARQ timers and CSI reporting becomes even more pronounced in this context, potentially requiring predictive scheduling mechanisms or modified HARQ cycles that account for variable propagation delays.

Furthermore, placing the RU on one satellite and DU/CU on another breaks the symmetry of deployment and complicates resource management. While Option 2a allows a one-to-one mapping of resource blocks and UE sessions to a single platform, Option 2b requires distributed session context management and reassembly of transport blocks over ISLs, which introduces complexity and increases error propagation risk.

Despite these challenges, Option 2b holds strategic promise in advanced NTN systems with extensive satellite constellations and sufficient processing redundancy. It aligns with the principles of disaggregated RAN and may allow for better resilience and resource optimization when paired with intelligent routing and synchronization frameworks. However, realizing such an architecture will require significant innovation in both protocol stack design and satellite networking mechanisms, potentially involving a space-adapted fronthaul standard or custom MAC/RLC implementations.


A practical variant of Option 2b involves placing both the RU and DU on one satellite, while hosting the CU on another satellite with which it communicates via an inter-satellite F1 interface. This configuration avoids the stringent latency and synchronization constraints of fronthaul protocols and instead leverages the more relaxed tolerances of the F1-C interface. According to 3GPP guidelines and industry experience, F1-C,  the control-plane interface between the DU and CU-CP,  can tolerate one-way propagation delays in the range of approximately 2 to 10 milliseconds \cite{Lar19, oran2025ntn}, depending on vendor implementation and buffering strategy.

Given that ISLs exhibit a typical one-way propagation delay of 
about 1.8 to 3.6 ms per hop, this allows for a maximum of %
2 to 3 
satellite hops between the DU and CU while remaining within acceptable latency budgets. 
This implies a proper ISL mesh planning, a centralized CU satellite could serve a cluster of multiple RU+DU nodes, each placed within a low-hop-radius region of the constellation.

This architectural design enables the pooling of CU functions, such as RRC signaling, bearer management, QoS enforcement, and mobility coordination, across several RU+DU access satellites, improving control-plane scalability and resource utilization. Moreover, centralized CU deployment simplifies handover coordination and potentially allows for tighter integration with onboard or edge-deployed RIC instances. However, the inherent mobility and variable topology of LEO constellations still present challenges for session anchoring, F1 association maintenance, and scheduling cycles. These may require adaptive transport-layer mechanisms (e.g., reliable routing overlays), dynamic F1 rebinding protocols, and robust session context management across satellite transitions. Nevertheless, this RU+DU - CU split offers a practical and scalable compromise, particularly for NTN deployments targeting mid-density access scenarios with centralized control logic and moderate beam steering capabilities.

\subsection{Option 3: Full gNB On-Board with UPF Functions}
\label{subsec:full_gnb_upf}


This option envisions a highly autonomous non-terrestrial node where the entire gNB (RU + DU + CU) and the \ac{UPF} are fully integrated on-board the satellite payload. A recent white paper from the O-RAN Alliance also notes that future versions (3GPP Rel-19) will consider scenarios that include “a gNB on the satellite [...] with the UPF on board” \cite{oran2025ntn}. This means the entire 5G NR base station and a 5G core user-plane node reside in space. An ESA-funded 5G study likewise assumed “a complete gNB on-board the satellite” to meet NTN latency constraints \cite{esa-5g-is-2024}. In practice, demonstration projects have already proposed putting full gNB functionality on satellites, with  ESA  explicitly showing “Full gNB on board with UPF might be also included \cite{esa-5g-spl-demo-2021}. This design supports local data breakout and reduced latency paths, particularly beneficial for edge-enabled services and delay-sensitive applications. 

Placing a UPF in the satellite means standard 5G core interfaces would traverse the satellite link. Notably, the N3 interface (gNB-CU user plane to UPF) becomes an internal link on the satellite, and the N4 interface (SMF to UPF) must operate over the satcom backhaul for control of the remote UPF. 3GPP’s ongoing work in Release 18 explicitly addresses this: the study on Support of Satellite Edge Computing via UPF on board was presented in the TSG SA WG2\#152E Electronic meeting \cite{3gpp-sa2-152e-2022}.  In essence, the network can deploy an onboard UPF as a local data anchor, so that traffic can break out directly at the satellite via its N6 interface to local applications or caches (enabling “satellite edge computing”).  The 3GPP study  solutions in \cite{3gpp-tr-23-700-27-2022} allow an onboard UPF to act as a local PDU Session Anchor, performing uplink classifier and branching for local routes.  The authors in \cite{jiang2023b5g} examine the UPF (and by extension the N3/N4 signalling) to be allocated on satellites, using the N9 interface between the satellite-based UPFs.  The paper in \cite{liu2024qos} demonstrate that satellite UPF deployment can reduce the energy consumption by 85.2\% while maintaining comparable
latency performance. At the same time, the paper in \cite{seeram6handover} find that cumulative conditional handover delay but demands 55\%–70\% more computational resources than the Split 7.2× architecture. Therefore, the full onboard gNB architecture was able to fulfil the strict requirements of ~50 ms delay for applications such as cloud gaming, while partial regenerative splits struggled to stay below 100 ms. This shows that low latency real-time services directly benefited from option 3. In addition, having the RAN and a portion of core  co-located at the edge (or on the satellite) can handle local mobility management and local data path offload aligning with  “ultra-flat, highly-distributed” architectures  \cite{esa-5g-is-2024}.

To summarise, the previous statements demonstrate that standard 5G core interfaces (N2/N3/N4) can be extended via satellite links to manage an on-board UPF, and that the UPF’s N6 interface can enable local breakout or edge computing applications in the satellite (e.g. content delivery or processing of data in the satellite itself).

\begin{figure}[htp!]
	\centering
    \includegraphics[width=0.45\textwidth] {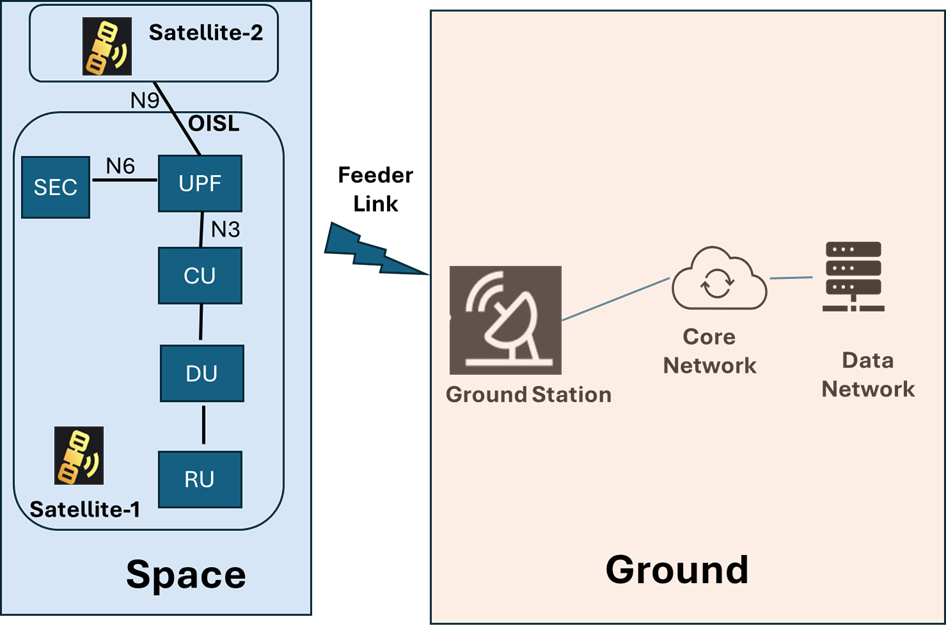}
	\caption{Option 3: Full gNB On-Board with UPF Functions}\label{fig:option3}
\end{figure}

As shown in the Fig. \ref{fig:option3}, the on-board gNB consists of the full stack: the RU handles analog-digital conversion and low PHY (IFFT/FFT, ADC/DAC), while the DU processes MAC, RLC, and high PHY functions such as modulation and coding. The CU manages both the control and user planes, interfacing with the on-board UPF over N3/N4 interfaces. Together, this enables complete gNB + UPF data path termination in space, decoupling the user plane from the terrestrial core for most of the traffic. A key advantage of this architecture is its compatibility with N9-based mesh routing via satellites with \glspl{ISL}. This allows packets to be forwarded between the satellites, reducing dependency on feeder links and enabling resilient and scalable routing across a satellite constellation. In addition, the presence of a fully operational UPF unlocks the N6 interface for integration with \ac{SEC} platforms. These SEC platforms can host applications at the data plane (e.g. \ac{CDN} caching, content filtering, analytics), enabling PDU-level processing in the satellite and service exposure. However, to realise this capability, significant on-board computing resources are required, e.g. dedicated CPUs and GPUs for packet routing, user processing and application execution.


Despite its advantages, this configuration also has significant disadvantages (i)   High computational demand necessitating powerful onboard hardware (e.g., CPUs, GPUs), payload complexity, and power consumption. These onboard processing units must be energy-efficient, radiation-tolerant and capable of handling significant data storage and a throughput of several Gbps. (ii)  Increased system complexity and power requirements.  Therefore, resource requirements and system complexity increase significantly, so thermal management, orchestration and fault tolerance are critical aspects of the design.



\subsubsection{Option 3a: The same satellite}

With regenerative satellites hosting UPFs, it becomes possible to form a mesh network in space. The satellites can route data traffic between each other at the user level, using ISLs as backhaul. In 5G terms, an N9 interface (the interface between two UPFs in the core network architecture) can be implemented via the inter-satellite link to route data between the UPFs on board the satellites \cite{jiang2023b5g}. With option 3a, both the complete gNB stack (RU, DU, CU) and the \ac{UPF}, including optional \ac{SEC} functions, are hosted on the same satellite platform. This co-location ensures that all interfaces — including N2, N3, N4 and N6 — remain within the same satellite, enabling extremely low latency between functions and simplified synchronisation between layers. The router, which is also on board, supports ISL-based N9 interfaces for routing user traffic across the satellites when required, but no signalling between the satellites is required for local gNB-UPF interaction. This design is advantageous for latency-sensitive applications, real-time analytics and mission-specific services that can run completely isolated from ground control (e.g. tactical IoT clusters or content delivery in remote areas).

From an implementation perspective, Option 3a simplifies interface management, eliminates feeder-link dependence for most user traffic, and provides a self-contained edge computing node in space. However, it requires a powerful satellite payload, as the integration of gNB, UPF, routing logic and SEC functions on a single platform requires significant processing overhead (CPU/GPU), memory and I/O throughput, which can increase payload weight and thermal complexity. This option is suitable for multifunctional satellites with high payload integration, especially in LEO constellations where short revisit times and proximity to the user terminal increase on-board autonomy.

\subsubsection{Option 3b: Different satellites}

Option 3b considers the gNB and UPF functions distributed across different satellites — for example, a satellite hosting the complete gNB stack offloads user plane traffic to another satellite equipped with UPF and optionally \ac{SEC} capabilities. Communication between these satellites takes place via ISL-based N9 interfaces, which enable dynamic user-plane routing in space. This setup supports functional disaggregation in the satellite domain and allows for specialisation of resources — some satellites focus on access (radio and RAN stack), while others are optimised for data processing, storage or application hosting.

This option allows for greater flexibility in the design of satellite constellations, as functions can be scaled independently and allocated dynamically based on mission requirements, geographic location of services or traffic load. For example, UPF-equipped satellites could serve as regional data anchors that aggregate and process traffic from multiple access satellites. However, this design introduces new challenges: it increases latency on the N3/N9 path between gNB and UPF due to inter-satellite forwarding, and it requires strict coordination and QoS control between the nodes. In addition, interruptions on the N9 path or satellite handovers could affect the continuity of the user plane unless mitigated by robust mobility anchoring and buffering. Option 3b is particularly relevant for mesh-enabled multi-orbit NTN architectures, where service chaining and distributed function placement are strategic enablers. It offers a modular and scalable path to support advanced satellite services, albeit with increased requirements for inter-satellite routing intelligence, network state propagation, and control-plane resilience.

\section{RIC placement options}
\label{sec:rics}


Deploying RICs in O-RAN-based NTNs requires addressing critical trade-offs between centralized intelligence and latency sensitivity. The placement of these components must consider several critical aspects, such as, the functional gNB splits described in previous section, the nature and origin of information exchanged by E2 nodes (i.e., CU and DU) and the capabilities offered by ISLs in terms of visibility and routing. While ground-based deployment simplifies management and AI/ML coordination, it introduces feeder-link delays that impair time-sensitive functions like beam management and handovers. Onboard deployment reduces latency but faces scalability and resource constraints in large constellations. With the above said, a ground-onboard tradeoff architecture is essential to effectively balance these requirements through strategic functional distribution. Accordingly, we propose different potential architectural extensions, one on top of the other, to the O-RAN framework in this section.


Regarding the information reported by E2 nodes, a DU provides near-real time metrics related with PRB usage, MAC scheduler statistics, Cell and UE-level KPIs, beam management info. In contrast, the CU mainly delivers control-plane events, including RRC setup, handovers, mobility-related events and user-plane statistics like throughput per QoS flow, flow-level delay/loss, session setup/teardown events. While both E2 nodes play a role, in practice, most near-RT data consumed by xApps on the near-RT RIC comes from the DU, leaving CU as a supplementary element.  In terms of communication infrastructure, current LEO satellite system commonly utilize up to four ISLs per satellite: two intra-plane links and two inter-plane links. Satellites within the same orbital plane maintain permanent visibility of one another, whereas inter-plane visibility varies depending on multiple factors such as orbital altitude, angular separation between orbital planes, and the field of view and targeting capability of the ISL terminals (laser or RF). In modern constellations like Starlink (550 km orbit height) or OneWeb (1,200 km orbit height), inter-plane links typically last between 5 and 15 minutes per satellite pair, and the link quality remains stable throughout the visibility window. Regarding routing capabilities of LEO constellation, current routing solutions employ predictive static routing tables. Based on known orbits, routes are generated to/from each region of the planet, anticipating how the network will move over time, so forwarding decisions are made on each satellite, like a router. Based on that, current satellite networks, like the one deployed by Starlink, can reach intercontinental distances in 6-10 hops, where the typical latency per hop is about 
2 to 4 ms.


\begin{figure*}[htbp!]
	\centering
    \includegraphics[width=0.8\textwidth] {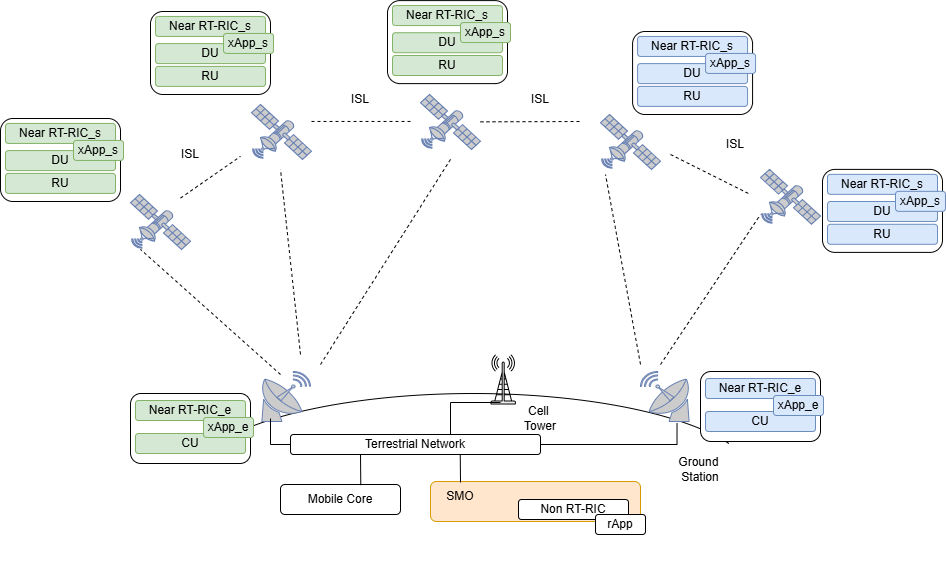}
	\caption{Architectural extension I: near-RT RIC split between Earth and Space}
    \label{fig:archI}
\end{figure*}

\subsection{Architecture Extension I: Splitting near-RT RIC Across Earth and Space}
\label{subsec:ricsI}


According to O-RAN specifications, among the requirements of the Non-RT RIC we found that it \textit{"shall support relevant AI/ML model training ... for generating/optimizing policies and intents to guide the behavior of applications in Near-RT RIC or RAN"}\footnote{https://docs.o-ran-sc.org/projects/o-ran-sc-nonrtric/en/amber/overview.html}. These operations demand significant computational resources and power consumption, making it practical to deploy the Non-RT RIC exclusively on the ground. Furthermore, its control-loop operates at a time scale exceeding one second, thus not imposing additional burdens and making feasible this location on ground to propagate policies through O1 and A1 interfaces. However, the propagation of these policies need to take into account the visibility of the implied elements (e.g., Near-RT RIC, DU, CU) based on their location and the availability of feeder links.

In contrast, the Near-RT RIC operates with a much tighter control-loop ranging from 10 ms to 1 s. This implies low-latency communication with E2 nodes, thus imposing great burden in their deployment location. LEO satellites introduce a 5 to 20 ms round trip delay, making the near-RT RIC placement on Earth, while the DU is in space, infeasible for strict near-RT use cases (e.g., scheduling optimization, HARQ, radio resource management). Thus, this advocates for a near-RT RIC in space, co-located with the DU, since most of near-RT data consumed by xApps comes from the DU, as stated before. According to the considered split in Section \ref{sec:splits}.\ref{subsec:du_space}, where CU is on Earth, we propose a similar logical split for the near-RT RIC, that is, splitting the near-RT RIC between Earth and space, as depicted in Figure~\ref{fig:archI}. In this split, we consider one component part of the near-RT RIC handling DU loops (on orbit) and one component handling CU loops (on Earth). Further digging, into this logical separation, lightweight inference xApps are deployed on satellites (e.g., for fast beam selection or HARQ prediction), while heavier models are hosted on earth. Such an approach could provide several architectural benefits, such as: i) keeping E2 interface local; ii) avoids long-latency cross-E2 links; iii) logical separation aligns with RAN function splits (DU vs. CU) and iv) allows hierarchical xApp design (i.e., local fast xApps (space) and global coordination or policy xApps (Earth)). This approach, however, introduces complexity and several new challenges. It necessitates the definition of a new dedicated interface to enable coordination between the space and Earth segments of the RIC. The current A1 interface would also require extension to facilitate communication between non-RT RIC and the different near-RT RIC to perform policy sharing. Additionally, consistent state management becomes more complex, particularly during mobility events where the DU movement triggers an inter-CU handover. Ensuring reliable and low-latency coordination in such cases requires robust state-consistency mechanisms and efficient context handover procedures. Finally, inter-segment communication must be secured against space-ground link vulnerabilities, such as jamming or spoofing, adding further complexity to the system design.



\subsection{Architecture Extension II: Network of near-RT RIC deployment in Space with dynamic E2 node assignment}
\label{subsec:ricsII}



\begin{figure*}[htp!]
	\centering
    \includegraphics[width=0.8\textwidth] {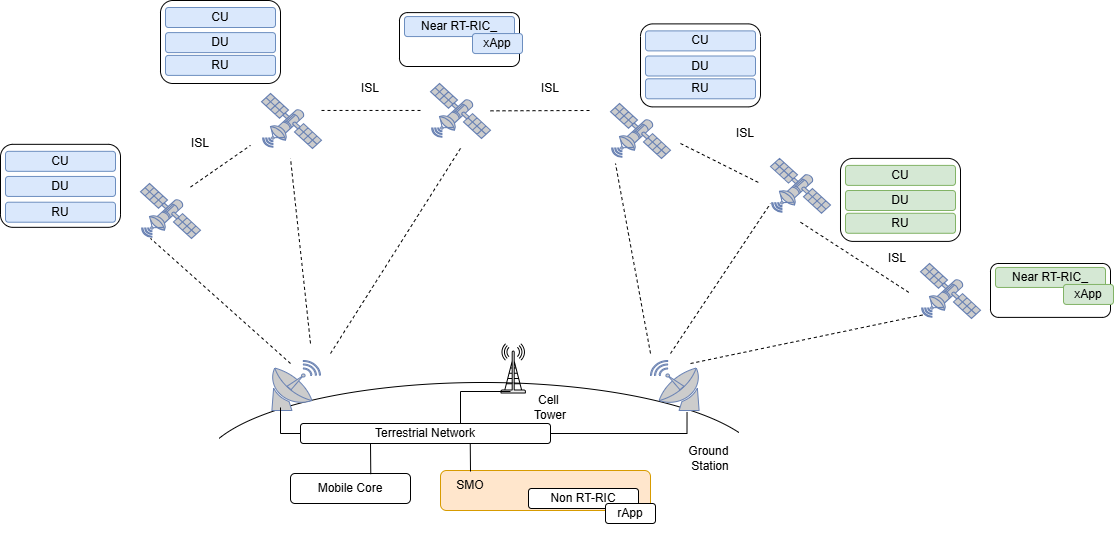}
	\caption{Architectural extension II: Network of near-RT RIC with dynamic assignment}
    \label{fig:archII}
\end{figure*}

When both the DU and CU are placed on orbiting satellites, as explored in Section \ref{sec:splits}.\ref{subsec:du_space} and Section \ref{sec:splits}.\ref{subsec:full_gnb_upf}, the whole near-RT RIC instance must be placed in space to meet tight control-loop timing requirements. However, as not highlighted in previous section, the addition of a new entity on the satellite is posing more burden to satellite capabilities in terms of processing and power consumption, thus requiring a very careful planning of the distribution of the different entities among satellites. In that sense, 
near-RT RICs instances will be deployed across the satellite constellation forming a network, where DU and CU attach to them dynamically based on their reachability, given the varying position, latency constraints and ISL availability, as depicted in Figure~\ref{fig:archII}. 

Within this scenario, the \ac{SMO} and/or the non-RT RIC, via O1 interface, could play a central role in orchestrating E2 node reassignment to maintain low-latency control loops and computational balance. By monitoring satellite trajectories and  using orbital knowledge (ephemeris) and real-time ISL status perfomance metrics, the system can predict when a given E2 node will exit the optimal control region of its current near-RT RIC instance. This foresight enables preemptive migration of the E2 interface connection to a better-positioned near-RT RIC instance. The reassignment is guided by multiple factors, including location, link quality, and load of satellite computational resources. 


This predictive handover strategy ensures that latency constraints for near-RT control loops are consistently met, while also distributing load computational tasks more evenly among satellites. Moreover, it enhances network resilience by reducing the risk of performance degradation due to congested nodes, deteriorating ISLs, or localized RIC failures. The approach also supports scalability, as new satellites equipped with RIC capabilities can seamlessly integrate into the distributed control plane.

Despite its potential, this architecture poses several challenges opening new research directions. A key limitation is the absence of standardized inter-RIC protocols in the current O-RAN framework, which hampers seamless coordination and state transfer between RIC instances. Realizing this architecture also requires predictive control mechanisms that can accurately forecast satellite movement and ISL status to initiate proactive and stateful handovers without service interruption. Such handovers can be even more complex if also considering the proposed extension in Section~\ref{sec:rics}.\ref{subsec:ricsI}, which requires proper associations between DU and CU instances and corresponding components of the near-RT RIC (i.e., earth vs space) if there are CU instances on Earth, as mentioned for Split 2. 
Managing consistent control state—including UE contexts, policy timers, and RAN-specific variables—across RICs under such tight time constraints introduces substantial synchronization overhead. Addressing these challenges is essential to unlock the full potential of near-RT RIC deployment in NTN environments.

\subsection{Architecture Extension III:  Splitting Non-RT RIC Across Earth and Space Distributed Clusters}

\begin{figure*}[htp!]
	\centering
\includegraphics[width=0.8\textwidth] {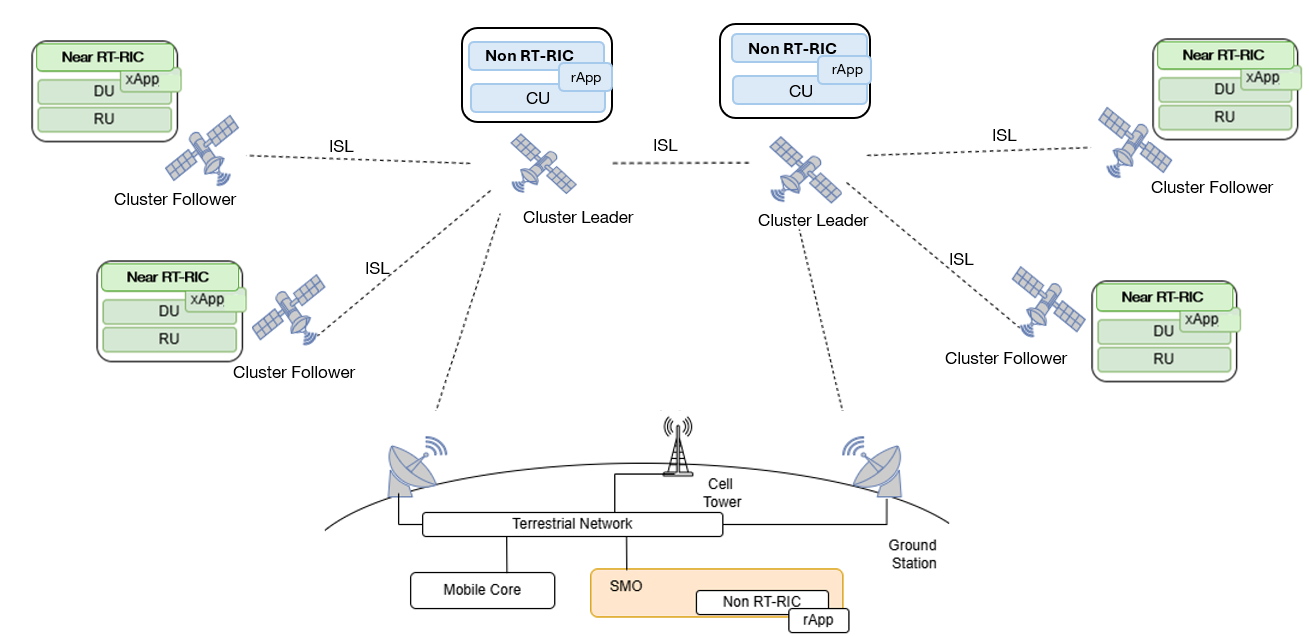}
	\caption{Architectural extension III: Splitting Non-RT RIC
Across Earth and Space Distributed Clusters.}
\label{fig:extension3}
\end{figure*}

In architecture extension I, we proposed partitioning the near-RT RIC between terrestrial and space segments, while retaining the non-RT RIC exclusively on the ground. As an alternative extension, we can consider a configuration where the non-RT RIC is distributed across both terrestrial and satellite clusters, with near-RT RIC functions migrated onboard satellites. This may also be interpreted as a variation of extension architecture II. Let us consider a hierarchical policy control framework as shown in  Fig. \ref{fig:extension3}. In this architecture, a ground-based non-RT RIC orchestrates network-wide policies, while distributed space-based RIC clusters enable localized optimization. Specifically, the satellite constellation is partitioned into multiple clusters, each managed by a designated cluster leader satellite equipped with a secondary non-RT RIC. The remaining satellites within each cluster, referred to as cluster followers, are responsible for executing near-RT RIC operations.

\textit{Level 1 (Ground Non-RT RIC):} This entity is responsible for long-term, network-wide orchestration, including AI/ML model training and global policy definition. By centralizing intelligence, the level 1 controller ensures coordinated resource allocation across satellite clusters and minimizes redundant computations. This ground non RT RIC can be envisaged to be located at \ac{SMO} for example.

\textit{Level 2 (Cluster Leader Non-RT RIC):} Each satellite cluster is managed by a leader satellite hosting a secondary non-RT RIC instance, effectively serving as an edge cloud node within the constellation. This layer refines and contextualizes global policies received from level 1, adapting them to cluster-specific operational conditions. Hierarchical ML frameworks are deployed at this level to decompose high-level objectives into actionable thresholds for follower satellites. Level 2 also manages the allocation and scaling of near-RT RIC instances across follower satellites, dynamically responding to real-time cluster demands. This edge autonomy mitigates dependence on ground connectivity, thereby ensuring resilience to temporary backhaul disruptions. The primary function of the cluster-level non-RT RIC is to perform localized optimization and distributed coordination within the cluster. (The precise definition and formation of satellite clusters remains an open research question.) This second non-RT RIC can be envisaged colocated with CU for the benefits explained previously. 

\textit{Level 3 (Follower Satellite Near-RT RIC):} At the lowest tier, colocated with DU and RU, the near-RT RIC instances operate directly on follower satellites, executing mission-critical tasks with sub-100 ms latency requirements. These entities are responsible for running xApps for real-time operations and implementing distributed learning algorithms, leveraging meta-policies provided by level 2 to optimize individual satellite performance. 

The hierarchical interaction among these layers enables multi-timescale optimization: ground-based systems perform strategic planning on hourly or daily timescales, while cluster-level RICs facilitate minute-level adaptation. This separation of concerns allows terrestrial controllers to focus on global network KPIs (e.g., aggregate energy consumption across constellations), whereas onboard controllers optimize local metrics (e.g., per-beam throughput or inter-satellite link latency). An additional advantage of this architecture is the facilitation of inter-satellite coordination, managed by the cluster-level non-RT RIC, which is particularly advantageous for scenarios involving heterogeneous satellite roles. Furthermore, the dynamic allocation of near-RT RIC instances by level 2 edge clouds supports seamless scalability as the constellation expands, thereby avoiding bottlenecks associated with centralized architectures.

Despite its potential, the proposed extension presents several challenges that merit further investigation. The optimal formation and definition of satellite clusters are complex tasks that require careful design to fully exploit the benefits of the level 2 non-RT RIC. In contrast to the current ORAN architecture, the proposed system necessitates new interfaces between ground-based and cluster-level non-RT RICs. Additionally, the dynamic association and dissociation of satellite clusters may pose significant challenges in mobility management, synchronization, and overall system robustness. For instance, the F1 interface may need to be extended to support dynamic reassignment as satellites move, enabling fast context transfer.  As in Extension I, enhancements to the A1 interface are also required to facilitate communication between the non-RT RIC and various near-RT RIC instances. Furthermore, the F1 interface may require protocol improvements to address variable ISL delay and jitter, as well as to support seamless handovers between moving satellites.

Finally, Table \ref{tab:o-ran_ntn_split}  summarizes the mapping with architectural splits and RIC placements.

\begin{table*}[htp!]
\centering
\scriptsize
\caption{Summary of O-RAN Based Split Design Options for Integrated NTN.}
\begin{tabular}{|p{1.6cm}|p{3.2cm}|p{2.8cm}|p{3.2cm}|p{3.5cm}|}
\hline
\textbf{Option} & \centering \textbf{O-RAN Functions\\in Space} & \centering \textbf{O-RAN Functions\\on Ground} & \textbf{Pros} & \textbf{Cons}  \\
\hline
\textbf{1a/1b} & 
\begin{itemize} [leftmargin=*]
\item \textbf{Option 1a:} RU + DU in the same satellite (Split 2) 
\item \textbf{Option 1b:} RU + DU in different satellites (class 1 + class 2)
\end{itemize} & \begin{itemize} [leftmargin=*]
\item \textit{Extension 1, 2 and 3:} CU + Near-RT RIC + Non-RT RIC + Core
\end{itemize}  & 
\begin{itemize}[leftmargin=*]
    \item Lightweight satellite
    \item Full reuse of ground infrastructure
    \item Lower Capex
\end{itemize} & 
\begin{itemize}[leftmargin=*]
    \item F1 interface needs to be implemented in feeder link.
    \item Satellite must be able to maintain 2 feeder links (for handover) 
    \item Need to support variable latency in the feeder link and seamless handovers
    \item High signalling load for CU handover
    \item In option 1b, new fronthaul technology is required
    \item High latency
    \item No local control
\end{itemize} \\

\hline
\textbf{2a/2b} & \begin{itemize} [leftmargin=*]
\item \textbf{Option 2a:} RU + DU + CU  in the same satellite 
\begin{itemize}
    \item \textit{Extension 2:} Near-RT with dynamic E2
    \item \textit{Extension 3:} Near-RT RIC in space 
\end{itemize}
\item \textbf{Option 2b:} DU, CU and RU in different satellites 
\begin{itemize}
    \item \textit{Extension 2:} Near-RT with dynamic E2
    \item \textit{Extension 3:} Near-RT RIC in space 
\end{itemize}
\end{itemize}  &  \begin{itemize} [leftmargin=*]
\item \textit{Extension 1:} Near-RT RIC splitting  in earth and space
\item \textit{Extension 2:} Non-RT RIC on earth
\item \textit{Extension 3:} Non-RT RIC splitting on earth and space
\end{itemize}   & 
\begin{itemize}[leftmargin=*]
    \item Reduce the dependence on earth -based controller and scheduler
    \item Remove the stringent timing requirements
    \item Reduce signaling burden
    \item Seamless UE handovers at gNodeB level
    \item Good compromise between latency and arch. complexity
    \item Easy to implement from protocol and hardware stand point
\end{itemize} & 
\begin{itemize}[leftmargin=*]
    \item Higher burden on the satellite on board processing unit
    \item Quasi-continuous feeder link visibility
    \item  No caching capabilities (increased latency and higher load feeder links)
\end{itemize} \\
\hline
\textbf{3a/3b} & \begin{itemize} [leftmargin=*]
\item \textbf{Option 3a:} RU + DU + CU + UPF in the same satellite 
\begin{itemize}
    \item \textit{Extension 2:} Near-RT with dynamic E2
    \item \textit{Extension 3:} Near-RT RIC in space 
\end{itemize}
\item \textbf{Option 3b:} DU, CU, RU and UPF in different satellites 
\begin{itemize}
    \item \textit{Extension 2:} Near-RT with dynamic E2
    \item \textit{Extension 3:} Near-RT RIC in space 
\end{itemize}
\end{itemize} &  \begin{itemize} [leftmargin=*]
\item \textit{Extension 1:} Near-RT RIC splitting  in earth and space
\item \textit{Extension 2:} Non-RT RIC on earth
\item \textit{Extension 3:} Non-RT RIC splitting on earth and space
\end{itemize}  &
\begin{itemize}[leftmargin=*]
    \item Reduced application latency
    \item Simplified interface management
    \item Elimination of feeder-link dependence
    \item Self-contained edge computing node in space
    \item Compatible with mesh connectivity through the N9 interface at the ISLs
\end{itemize} &
\begin{itemize}[leftmargin=*]
    \item High computational depends on onboard hardware
    \item Higher payload complexity and power consumption
    \item Interruptions on the N9 path or satellite handovers affects the continuity of the user place
\end{itemize}  \\
\hline
\end{tabular}
\label{tab:o-ran_ntn_split}
\end{table*}

\section{Discussions and Future Work}
\label{sec:discussion}


\subsection{Open Challenges and Gaps Identified}



Disaggregating gNB components in NTNs implies novel architectural and operational challenges due to the mobility of hosting platforms like LEO or MEO satellites and the strict latency and synchronization constraints of terrestrial RAN interfaces. While functional split architectures (presented Options 1 to 3 in Section~\ref{sec:splits}) offer deployment flexibility, they require reevaluating all core interfaces and control functions in the 5G stack.

In Option 1, the DU and RU are co-located onboard the satellite, while the CU remains on the ground. This reflects a Split 2 architecture, where lower-layer functions (e.g., PHY, MAC, HARQ) are processed in space, reducing latency for time-sensitive operations. However, the F1 interface must traverse the feeder link between satellite and Earth. This link is sensitive to propagation delay, intermittency, and jitter, which may compromise control-plane responsiveness and degrade user-plane throughput under variable link quality. In contrast, N2 and N3 interfaces (connecting the CU to the 5G Core) and the E2 interface (between CU and near-RT RIC) remain fully terrestrial and are unaffected by the satellite hop. Still, the end-to-end responsiveness of the RAN is limited by the F1 bottleneck, particularly during handovers or real-time reconfigurations initiated from the CU.

Option 2a places the RU, DU, and CU on a single satellite. This removes the need for internal F1 and E1 links and avoids fronthaul timing issues, but introduces feeder link constraints for N2/N3. If the RIC stays grounded, the E2 interface suffers from propagation delay, impacting time-sensitive functions like beamforming or mobility control. Deploying near-RT RICs onboard may mitigate these delays but brings added complexity for lifecycle management and reliability under satellite resource constraints. Option 2b spreads RU, DU, and CU across different satellites connected by ISLs. This modular approach supports scalability and redundancy, but stresses protocol limits, for instance in F1-C interface. 
This limits CU placement and requires strict control over constellation topology. E1 separation between CU-CP and CU-UP faces similar issues with added complexity for session state consistency across moving platforms. The Xn interface, essential for inter-gNB coordination, is poorly suited for dynamic satellite environments and needs redesign for beam-aware handovers and fast UE context forwarding. E2 becomes critical in this distributed setup. Near-RT control suffers under multi-hop signaling delays and losses. Solutions include embedding near-RT RICs with each DU or CU or clustering RICs across local satellites. These strategies require efficient synchronization and orchestration. xApps must be space-aware, using telemetry, visibility maps, and link conditions to manage HARQ, beamforming, and load balancing with limited compute resources. As satellites move or become unavailable, DU and CU nodes may need to migrate. Existing context transfer protocols are insufficient for maintaining continuity across F1 and E1 under dynamic topologies. Synchronization, vital for RAN performance, is also challenged by variable ISL delays. New timing methods or predictive mechanisms are needed to avoid HARQ and scheduling issues from skew and misalignment.

Option 3 adds onboard UPF, turning satellites into edge data nodes with local breakout over N6. This benefits MEC use cases and eases backhaul load but creates new problems for routing, security, and state persistence. First of all, hosting gNB and UPF functions on a satellite imposes significant processing demands on the spacecraft. The satellite must perform real-time baseband processing, packet routing, and possibly edge computations within strict size, weight, power, and radiation constraints. The ESA 5G Space-Based Infrastructure paper\cite{esa-5g-is-2024} estimates the power needed: in a mid-term LEO scenario, just the NR baseband processing (full gNB) plus feeder link modem could draw on the order of 130+ watts on board where the user link (on-board NR baseband processing/full gNB) consumes approximately 78.6 W and the feeder link modem (DVB‑S2X) draws about 55.9 W. The computing load is significantly higher –for a fully orbital gNB compared to split architectures (55-70 \% more demands) \cite{seeram6handover}. This means that more powerful onboard CPUs/DSPs or dedicated accelerators are required. In fact, researchers are also actively researching specialised hardware for satellite edge processing. The article in \cite{benelli2024gpu}, for example, presents “GPU@SAT”, a GPU-like high-performance accelerator implemented on an FPGA and designed for on-board processing of intensive tasks in space. These efforts emphasise the need for space-grade computing hardware (multi-core CPUs, FPGAs/GPUs, AI accelerators) for the 5G RAN and the core functions in orbit. At the same time, UPF failure or satellite loss also demand fast recovery and session relocation. For this reason, RICs must now account for N6 optimization and service placement, increasing their role in system orchestration.

To summarise, there are still unresolved gaps in all options. No current interface protocol supports the mobility of hosting nodes. RIC placement and xApp migration are not standardised for space environments.  RICs lack awareness of ISL conditions and there is no scalable satellite-native synchronization scheme. For gateways and session anchors, ground contact is still critical, making the systems vulnerable to link outages. Finally, on-board processing must operate under tight power limits, requiring lightweight and resilient designs. For this reason, overcoming these challenges requires more than the extension of terrestrial standards. They require new protocols, timing architectures and control strategies tailored to dynamic, delay-sensitive and distributed satellite networks.

\subsection{Future Directions}


As NTN integration progresses toward commercialization in 5G-Advanced and 6G ecosystems, several future directions must be pursued to mature the architectural and functional split models proposed in this work. Future research can address  real-world implementation aspects such as synchronization across inter-satellite links, AI-native RIC orchestration under dynamic connectivity, and security implications of distributed control planes. First, rather than relying on static placements of DU, CU, and UPF functions, future systems should support advanced reconfigurable splits that adapt to link quality, mobility, and computational load. This will require highly sophisticated orchestration platforms capable of real-time telemetry ingestion, predictive resource management, and low-latency control delegation across space-ground domains. Second, the interplay between Near-RT and Non-RT RICs—whether located on Earth, in LEO nodes, or across ISLs—needs to be optimized for both global policy enforcement and local adaptation. Federated learning, semantic policy distillation, and intent-based networking will play a central role in managing these disaggregated control loops. Third, seamless support for Split 2 (CU/DU) 
and Split 7.2x (high/low PHY) across heterogeneous vendor environments and satellite platforms requires enhancements to the O-RAN Alliance’s Open Fronthaul and F1/W1 interfaces to accommodate space-specific constraints such as longer round-trip delays and Doppler shifts. Additionally, efforts toward aligning these architectures with evolving 3GPP and O-RAN Alliance specifications will be essential to ensure standard-compliant and interoperable NTN deployments.

\section{Conclusions}
\label{sec:conclusions}

This paper presented a structured framework for analyzing and designing architectural and functional split options in integrated terrestrial and non-terrestrial networks. By leveraging the modularity of the O-RAN architecture, we evaluated multiple configurations for distributing RAN functions (RU, DU, CU) and UPF components across space and ground domains. Our analysis identified key trade-offs related to latency, onboard complexity, feeder link dependency, and autonomy, culminating in three principal design options: (i) DU in space, (ii) full gNB in space, and (iii) full gNB with UPF and SEC functions onboard. For each option, sub-variants were considered depending on whether the functions were co-located on the same satellite or distributed across multiple nodes connected via \glspl{ISL}.

We also proposed RIC placement strategies to complement these split designs, outlining both ground-based and hybrid control architectures. The mapping between split configurations and RIC distribution showed that control plane responsiveness and optimization capabilities depend heavily on the persistence and latency of inter-domain links. Overall, we highlight that increasing autonomy and lowering latency generally requires moving more intelligence onboard, but this comes with greater power, complexity, and implementation costs. Each option must be evaluated based on mission goals, satellite capabilities, and the required level of ground independence. The findings of this work suggest that function placement in NTNs must consider not only traditional performance metrics, but also operational and regulatory constraints unique to satellite environments.

\balance

\bibliographystyle{ieeetr}
\bibliography{biblio}

\begin{IEEEbiography}
[{\includegraphics[width=1in,height=1.25in,clip,keepaspectratio]{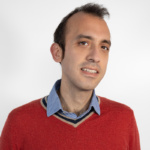}}]{JORGE BARANDA} (Senior, IEEE) received the M.S. and Ph.D. degrees in electrical engineering from the Technical University of Catalonia, in 2008 and 2021, respectively. He has participated in 15+ projects funded by European Commission, and Spanish Ministry of Science. He is currently a Senior Researcher with the Services as Networks (SaS) Research Unit, CTTC, Barcelona, Spain. His current research interests mainly include the management and orchestration of SDN/NFV 5G/6G mobile networks including multi-administrative domain aspects and the integration of heterogeneous technological domains like TN-NTN.
\end{IEEEbiography}

\begin{IEEEbiography}[{\includegraphics[width=1in,height=1.25in,clip,keepaspectratio]{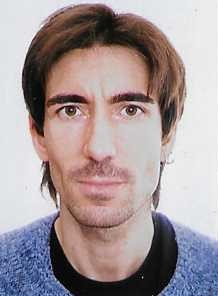}}]{MARIUS CAUS} (Senior, IEEE) received the M.Sc. and Ph.D. degrees in telecommunications engineering from the Universitat Politècnica de Catalunya (UPC), Barcelona, Spain, in July 2008 and December 2013, respectively. In 2018, he received the two-year postdoctoral Juan de la Cierva Fellowship from the Spanish Government. He is currently a Researcher with CTTC. He has participated in several projects funded by European Commission, European Space Agency, and Spanish Ministry of Science. His main research interests include filter bank-based multicarrier systems, signal processing for communications, and satellite communications.
\end{IEEEbiography}

\begin{IEEEbiography}
[{\includegraphics[width=1in,height=1.25in,clip,keepaspectratio]{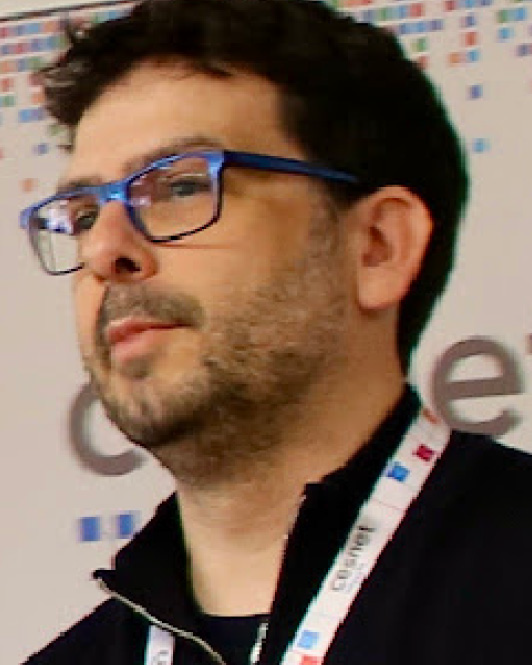}}]{LUIS BLANCO} received the Postgraduate degree in data science and quantitative finance, and the first
M.Sc. and Ph.D. degrees in Telecommunications engineering from the Polytechnic University of Catalonia, along with the second M.Sc. and MBA
degrees in research from ICT (MERIT).
Dr. Blanco is a Senior Researcher with CTTC. He has actively participated in more than 25+ research projects funded by public and private entities, including EC and ESA-funded projects,
national and industrial projects. Furthermore, he has also led multiple projects with European industry. His research interests focus on AI and optimization for wireless communications in beyond 5G and 6G and TN-NTN integrated systems, with special emphasis in the RAN.
\end{IEEEbiography}

\begin{IEEEbiography}
[{\includegraphics[width=1in,height=1.25in,clip,keepaspectratio]{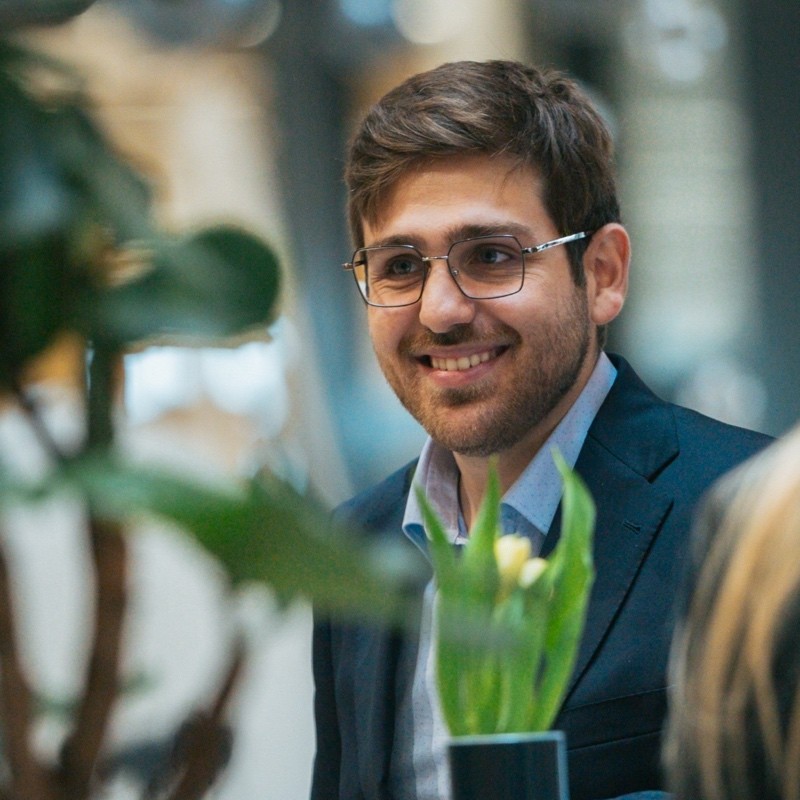}}]{CRISTIAN J. VACA-RUBIO}
 (Member, IEEE) is currently a Researcher at the Centre Tecnològic de Telecommunicacions de Catalunya (CTTC) in Barcelona, Spain. He received the Telematics engineering degree, and the master’s degree in Telematics and Telecommunication Networks from University of Malaga (UMA), Malaga, Spain, in 2018, and 2019, respectively. He received the Ph.D. degree in Wireless Communications from Aalborg University, Aalborg, Denmark in 2023. In parallel to his telematics and master studies, he worked as a research assistant with the Communication Engineering Department at University of Malaga from February 2018 until July 2019 in a joint project with DEKRA Gmbh. He pursued the Ph.D. degree with the Connectivity Section, Electronic Systems Department, Aalborg University, under a Marie Curie Fellowship as an Early Stage Researcher (ESR) in the H2020 ITN Wind Mill Project, where he joined in September 2019. During his Ph.D., he was a visiting researcher at Mitsubishi Electric Research Laboratories (MERL), Cambridge, USA for a period of 9 months. His main research activities are in wireless communications and sensing, machine learning, massive MIMO, mobile networks, telematics, and applied statistics in video streaming. He received the IEEE International Conference on Acoustics, Speech and Signal Processing top 3\% paper award in 2023. He is also the author of a  patent along with MERL co-authors.
\end{IEEEbiography}

\begin{IEEEbiography}
[{\includegraphics[width=1in,height=1.25in,clip,keepaspectratio]{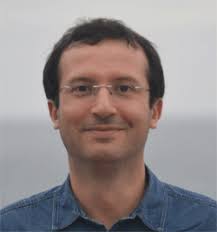}}]{ENGIN ZEYDAN} (Senior, IEEE)
received the B.S. and M.S. degrees from the Department of Electrical and Electronics Engineering, Middle East Technical University, Ankara, Türkiye, in 2004 and 2006, respectively, and the Ph.D. degree from the Department of Electrical and Computer Engineerin, Stevens Institute of Technology, Hoboken, NJ, USA, in 2011. He is a Senior Researcher with the Services as networks, Centre Tecnologic de Telecomunicacions de Catalunya (CTTC), Barcelona, Spain. Before joining CTTC in 2018, he worked as an R\&D Engineer for Avea (a Turkish mobile operator) from 2011 to 2016, a Senior R\&D Engineer in Turk Telekomunikasyon A.S from 2016 to 2018, and a Part-Time Instruction with the Electrical
and Electronics Engineering Department, Ozyegin University from January 2015 to June 2018. He has been primarily responsible for carrying out European Commission and nationally funded research activities at CTTC, Türk Telekomunikasyon, Avea. He is currently the Project Coordinator of the Horizon Europe UNITY-6G European Project since January 2025. He was the Project Coordinator of the H2020 MonB5G European Project from November 2021 to April 2023. He has also been involved in other European level projects such as 5Growth from 2019 to 2022 and Clear5G (as WP leader) from 2017 to 2018, FP7 projects MOTO (as WP leader) from 2012 to 2015, and CROWD from 2014 to 2015 in
collaborations with various industries and universities. He is a co-author of over 150+ papers in international journals and conferences and 12 patents (11 granted in Turkish Patent Institute and one granted under European Patent Office). His research interests are in the areas of telecommunications,
data engineering/science and network security.
\end{IEEEbiography}

\begin{IEEEbiography}
[{\includegraphics[width=1in,height=1.25in,clip,keepaspectratio]{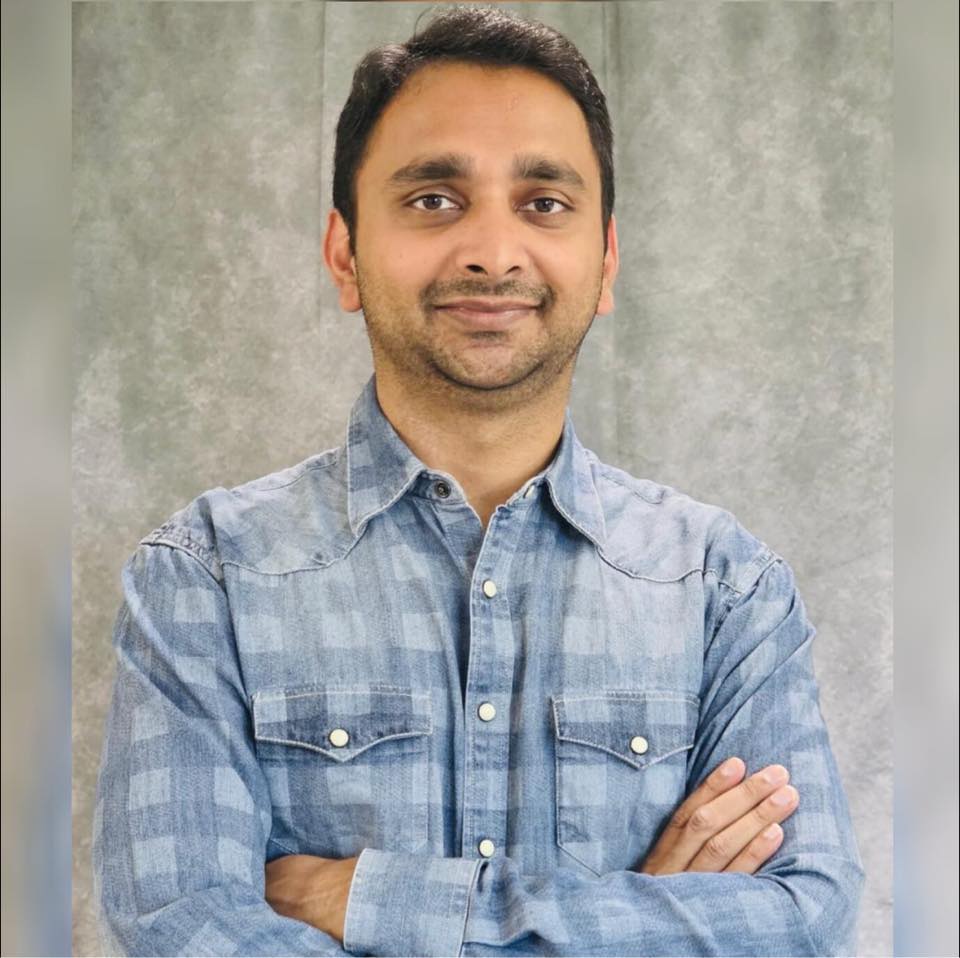}}]{KAPAL DEV} (Senior, IEEE) is an Assistant Professor in the Department of Computer Science at Munster Technological University (MTU), Ireland. He was awarded the PhD degree by Politecnico di Milano, Italy in 2019 under the prestigious fellowship of Erasmus Mundus funded by the European Commission. He is recognized with numerous awards, including the IEEE ComSoc EMEA Outstanding Young Researcher (2022) and the Tom Brazil Excellence in Research Award (2023). He is among the top 45 scientists elected to the Global Young Academy (GYA) 2024 and has secured over €2.5 million in competitive funding as a Principal Investigator. Dr. Dev holds editorial roles as associate editor as well as guest editor in major IEEE, Elsevier Q1 journals and magazines, NATURE, and other renowned publishers, He has led workshops at top-tier conferences such as ACM MobiCom, IEEE Globecom, and IEEE Blockchain, and his research on data and model security and privacy, Agentic AI, and Industry 5.0 has produced over 100 publications in leading scientific venues. In addition, he has supervised more than 25 postgraduate students and serves as an external evaluator for high-profile funding bodies. He is a Senior Member of the IEEE and a professional member of ACM, underlining his commitment to the global research community. 
\end{IEEEbiography}

\begin{IEEEbiography}
[{\includegraphics[width=1in,height=1.25in,clip,keepaspectratio]{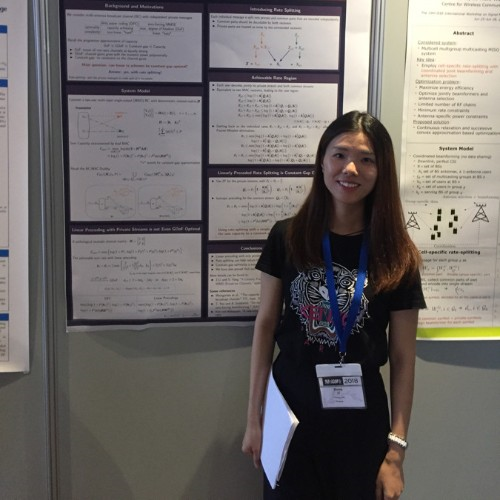}}]{ZHENG LI} (Member, IEEE) received her B.E. degree in Information Engineering from Southeast University, China, in 2013. She continued her studies and received double M.S. degrees in Telecommunications —one from CentraleSupélec, France, in 2015, and another from Southeast University, China, in 2016. She received her PhD degree in Telecommunications from CentraleSupélec, France, in 2020. Since 2020, she has been a researcher at Orange Innovation, France. Since 2021, she is the Orange delegate in ORAN WG4 on Open FH. Since January 2025, she serves as the Orange delegate in 3GPP RAN3. Her current research interests focus on the application of AI/ML in future networks, as well as the integration of TN and NTN.
\end{IEEEbiography}

\begin{IEEEbiography}[{\includegraphics[width=1in,height=1.25in,clip,keepaspectratio]{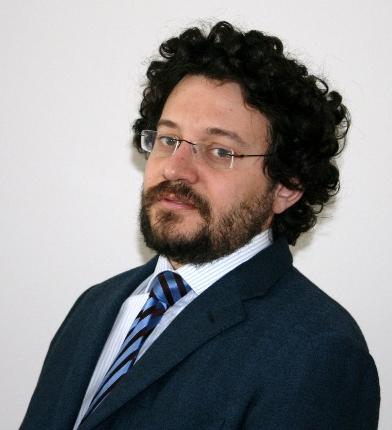}}]{TOMASO DE COLA}  (Member, IEEE) received the "Laurea" degree (with honors) in telecommunication engineering, in 2001, the Qualification degree as a Professional Engineer in 2002, and the Ph. D. degree in Electronic and Computer Engineering, Robotics, and Telecommunications in 2010 from the University of Genoa, Italy. Since 2008, he has been with the German Aerospace Centre (DLR), where he has been involved in different European Projects focusing on various aspects of DVB standards, CCSDS protocols, and testbed design. He is a co-author of more than 100 papers, including international conferences and journals. He was the chair of the Satellite and Space Communications (SSC) technical committee within ComSoc and of the SatCom working group within the ETP NetworldEurope platform. Currently he serves as area director of the Space-Internetworking Services (SIS) within the CCSDS standardization body. He has been involved in many editorial initiatives in IEEE journals and magazine as guest and associate editor.
\end{IEEEbiography}

\end{document}